\documentclass[reprint, amsmath,amssymb,aps, onecolumn]{revtex4-2}

\usepackage{graphicx}
\usepackage{subfigure}
\usepackage{subfloat}
\usepackage{amssymb}
\usepackage{amsmath}
\usepackage{multirow}
\usepackage{dcolumn}
\usepackage{bm}
\usepackage{wasysym}
\usepackage{cancel}

\usepackage{txfonts}
\usepackage{lineno}

\begin{document}

\title{Effect of Turbulent Kinetic Helicity on Diffusive $\beta$ effect for Large Scale Dynamo}

\author{Kiwan Park}
 \affiliation{Soongsil University, 369, Sangdo-ro, Dongjak-gu, Seoul 06978 Republic of Korea\\ pkiwan@ssu.ac.kr\\}

\date{\today}

\begin{abstract}
We investigated a plasma system with kinematic viscosity \(\nu = 0.006\) and magnetic diffusivity \(\eta = 0.006\), driven by helical kinetic energy, to study the dynamics of energy and helicity in magnetic diffusion. Using the numerical data obtained, we explored methods to determine the \(\alpha\) and \(\beta\) coefficients that linearize the nonlinear electromotive force (EMF) and the dynamo process. Initially, we applied conventional statistical approaches such as mean field theory (MFT), direct interaction approximation (DIA), and eddy-damped quasinormal Markovian (EDQNM) closure. We then proposed a simpler alternative method using large-scale magnetic data and turbulent kinetic data to calculate \(\alpha\) and \(\beta\). Our findings show that while \(\alpha\) qualitatively aligns with theoretical predictions, \(\beta\) remains negative, indicating an inverse cascade of energy through magnetic diffusion. This deviates from conventional models and was further analyzed using a recursive method in the second moment identity, revealing that small-scale kinetic helicity couples with large-scale current density to transport energy inversely. We validated our method by reproducing the numerically calculated data. The consistency between our method and direct numerical simulations (DNS) suggests that the negative diffusion process in plasma has a physical basis.
\end{abstract}

\maketitle

\section{Introduction}
Most celestial plasma systems are significantly influenced by magnetic fields, which play a crucial role in their dynamics. Magnetic fields extract energy from turbulent plasma, and the amplified fields react back on the system, constraining its evolution. On a macroscopic scale, magnetic fields are known to regulate the rate of star formation and the development of accretion disks \citep{1991ApJ...376..214B, 2005MNRAS.362..369M}. Additionally, the balance between magnetic pressure and plasma pressure can determine the stability of the system, as seen in phenomena such as the sausage, kink, or Kruskal-Schwarzschild instabilities \citep{2003Phpl.book.....B}. Furthermore, the ubiquitous magnetic fields, having existed since just after the Big Bang, may have also played a critical role in nucleosynthesis, the production of matter in the Universe. Perturbed electrons, influenced by the Lorentz force, increase electron density through superposition, reducing the potential barrier between interacting nuclei and facilitating fusion \citep{2023arXiv230713373P, 2024PhRvD.109j3002P}. Magnetic fields are a nearly unique physical entity capable of influencing a plasma system while maintaining electrical neutrality.\\

According to Maxwell's theory, magnetic fields in free space propagate infinitely while gradually decreasing in amplitude or field density. However, in plasma, where massive charged particles interact with the magnetic field, this propagation requires significantly more energy to overcome the interference from these heavy particles. Often, magnetic eddies, combined with particle motion, expand by reducing their eddy scales alongside fluid eddies. From the perspective of electromagnetism, this behavior contrasts with the intrinsic nature of magnetic fields. For magnetic eddies to grow in scale, specific conditions must be met to overcome the eddy turnover time or the inertia of massive particle eddies.\\

The amplification of magnetic fields migrating toward either larger or smaller scales is referred to as the dynamo. This process involves the conversion of kinetic energy into magnetic energy and its subsequent transport within the plasma system. The converted \(B\)-field cascades toward either the large-scale or small-scale regime, both processes involving the induction of the \(B\)-field through electromotive force (EMF, \((\mathbf{U} \times \mathbf{B})\), where \(\mathbf{U}\) represents fluid velocity). The migration of magnetic energy toward the large-scale regime is known as an 'inverse cascade,' resulting in a large-scale dynamo (LSD). In contrast, the 'cascade of energy' refers to a small-scale dynamo (SSD). The cascade of energy to smaller scales is commonly observed in hydrodynamics (HD) and magnetohydrodynamics (MHD). However, the inverse cascade of energy requires more stringent conditions, driven by factors such as helicity, differential rotation \citep{1966ZNatA..21.1285S, 1978mfge.book.....M, 1980opp..bookR....K, 2001ApJ...550..824B}, or magnetorotational instability (MRI), also known as the Balbus-Hawley instability \citep{1991ApJ...376..214B}. In SSD, non-helical magnetic energy (\(E_M\)) cascades toward smaller scales, resulting in the peak of \(E_M\) forming between the injection and dissipation scales \citep{2005PhR...417....1B}. If the growth rate of the magnetic field depends on magnetic resistivity, the process is called a slow dynamo; otherwise, it is known as a fast dynamo. The migration and amplification of the magnetic field are influenced by various factors, and the critical conditions for these processes remain an ongoing topic of debate.\\

In the presence of a helical magnetic field (\(\nabla \times \mathbf{B} = \lambda \mathbf{B}\)), the turbulent electromotive force \(\langle \mathbf{u} \times \mathbf{b} \rangle\) is known to be expressed as \(\alpha\overline{\mathbf{B}} - \beta \nabla \times \overline{\mathbf{B}}\), regardless of whether the turbulence is driven by kinetic or magnetic forces. Determining the \(\alpha\) and \(\beta\) coefficients is crucial for explaining the evolution of the solar magnetic fields and other astrophysical ones. For instance, the magnetic induction equation reveals that the evolutions of poloidal and toroidal magnetic fields are constrained by the \(\alpha\) coefficient \citep{1982GAM....21.....P, 2014ARA&A..52..251C}. A precise understanding of these coefficients is valuable for forecasting space weather, which has significant implications for Earth's climate.\\

There have been efforts to calculate these coefficients \citep{1976JFM....75..657K, 2002GApFD..96..319B, 2024MNRAS.530.3964B}. Currently, only approximate forms of the \(\alpha\) and \(\beta\) coefficients can be obtained through dynamo theories such as mean field theory (MFT), the eddy-damped quasi-normal Markovian (EDQNM) approximation, or the direct interaction approximation (DIA). These theories generally suggest that \(\alpha\) is related to residual helicity \(\langle \mathbf{b} \cdot (\nabla \times \mathbf{b}) \rangle - \langle \mathbf{u} \cdot (\nabla \times \mathbf{u}) \rangle\). The \(\beta\) coefficient, on the other hand, is related to turbulent energy \(\langle u^2 \rangle + \langle b^2 \rangle\).\\

Additional theoretical and experimental work has been conducted on negative magnetic diffusivity \citep{1999GApFD..91..131L}, and references therein. These studies are based on \(\alpha - \alpha\) correlations in strong helical systems. Additionally, \cite{2014NJPh...16g3034G} experimentally found that turbulent magnetic diffusivity \(\eta_{\text{turb}}\) was negative. They argued that the net diffusivity \(\eta_{\text{turb}} + \eta\) became positive again. However, we believe that the overall dynamo effect weakened in their work. Negative magnetic diffusivity was also observed in another liquid sodium experiment \citep{2014PhRvL.113r4501C}, where it was found that small-scale turbulent fluctuations (\(\sim u\)) contribute to the negative magnetic diffusivity in the interior region. Recently, \cite{2024MNRAS.530.3964B} suggested an iterative removal of sources (IROS) method that uses the time series of the mean magnetic field and current as inputs. This approach is quite mathematical, producing detailed components of \(\alpha_{ij}\) and \(\beta_{ij}\). We also investigate the effect of magnetic diffusion \(\beta\) on the helical large-scale dynamo. We also aim to make this theoretically refined approach applicable to experiments or observations.\\



In Section 2, we describe the numerical model and code. Chapter 3 presents the results of numerical calculations, while Chapter 4 explains the related theory. Some of these theories have been introduced in our previous work (\cite{2023ApJ...944....2P}< references therein), but for consistency and readability, they are explained again in more detail with additional content. We also include some IDL scripts used to create plots from the numerical data. The final chapter provides a summary. The most important feature of this paper is its use of large-scale magnetic energy and magnetic helicity data, which are relatively easy to measure, to apply theoretical methods for obtaining \(\alpha\) and \(\beta\) coefficients. This approach is then verified using turbulent kinetic data and is employed to directly reproduce the evolution of the large-scale magnetic field obtained numerically for comparison.

\begin{figure*}
    {
   \subfigure[$U_{rms}$ \& $B_{rms}$]{
     \includegraphics[width=9.0 cm]{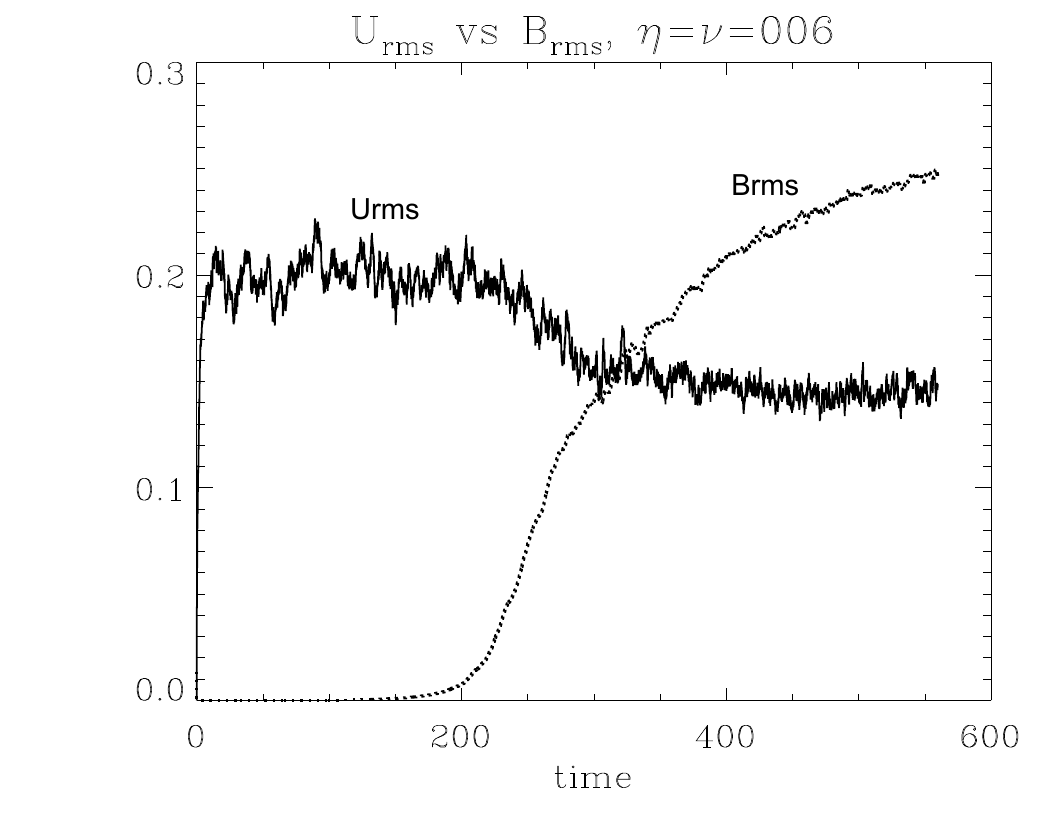}
     \label{f0E}
    }\hspace{-13 mm}
   \subfigure[$\langle k\rangle_{kin}$ \& $\langle k\rangle_{mag}$ ]{
   \includegraphics[width=9.2 cm]{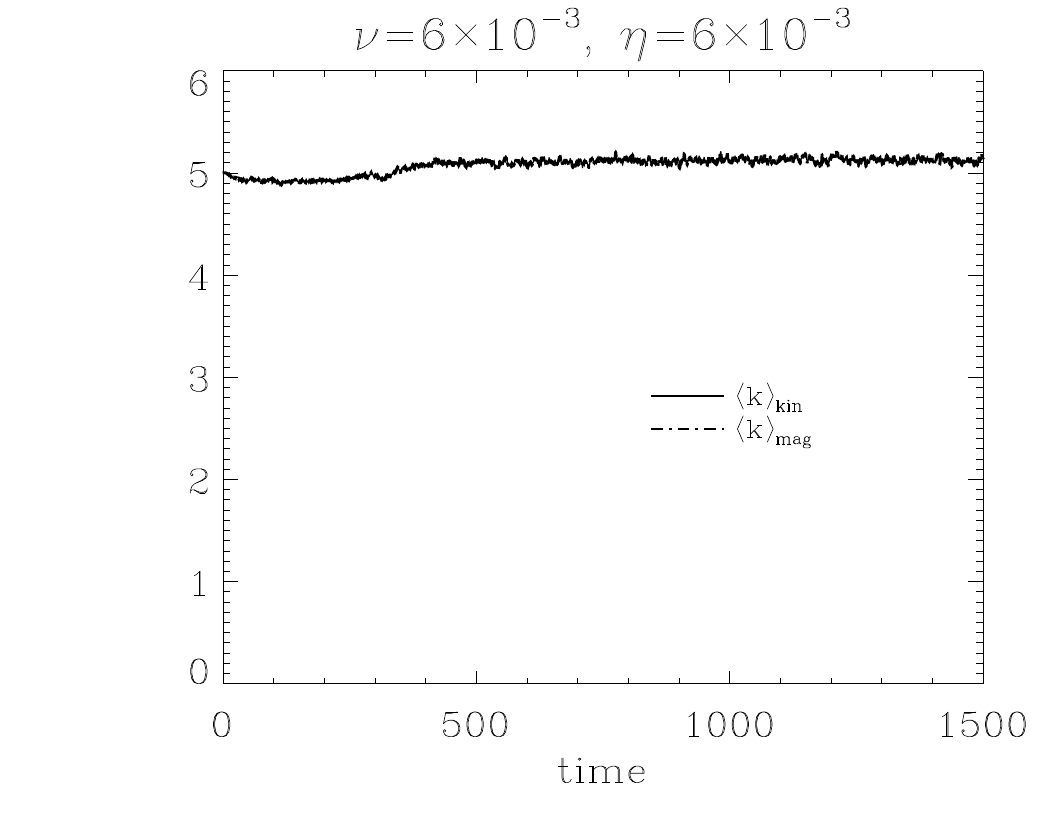}
     \label{f0k}
   }
}
\caption{(a) The mean velocity and magnetic field were calculated using the following definitions:
$U_{rms} = \big(2\int E_V dk\big)^{1/2}$, $\,B_{rms} = \big(2\int E_M dk \big)^{1/2}$.
(b) $\langle k\rangle_{kin} = \int kE_V dk/\int E_V dk$, $\langle k \rangle_{mag} = \int kE_M dk/\int E_M dk$. The inverse cascade in LSD does not imply that the entire energy migrates to the large scales. Instead, $E_M$ in the large scale regime at $k=1$ exceeds the externally applied forcing energy $E_V$ at $k=5$.}
\end{figure*}

\begin{figure*}
    {
   \subfigure[$\overline{U}^2$, $\overline{B}^2$, $\alpha$, $\beta$]{
     \includegraphics[width=9.2 cm]{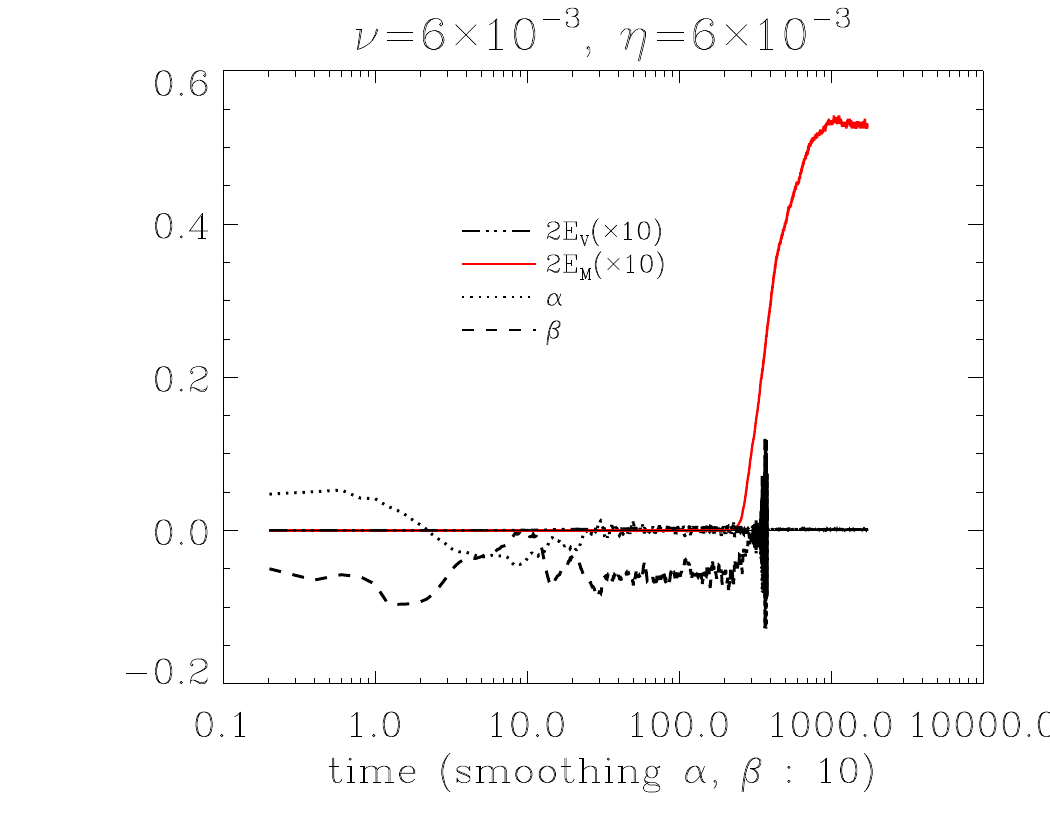}
     \label{f1}
    }\hspace{-13 mm}
   \subfigure[$|H_V(k)|$, $2E_V(k)$]{
   \includegraphics[width=9.2 cm]{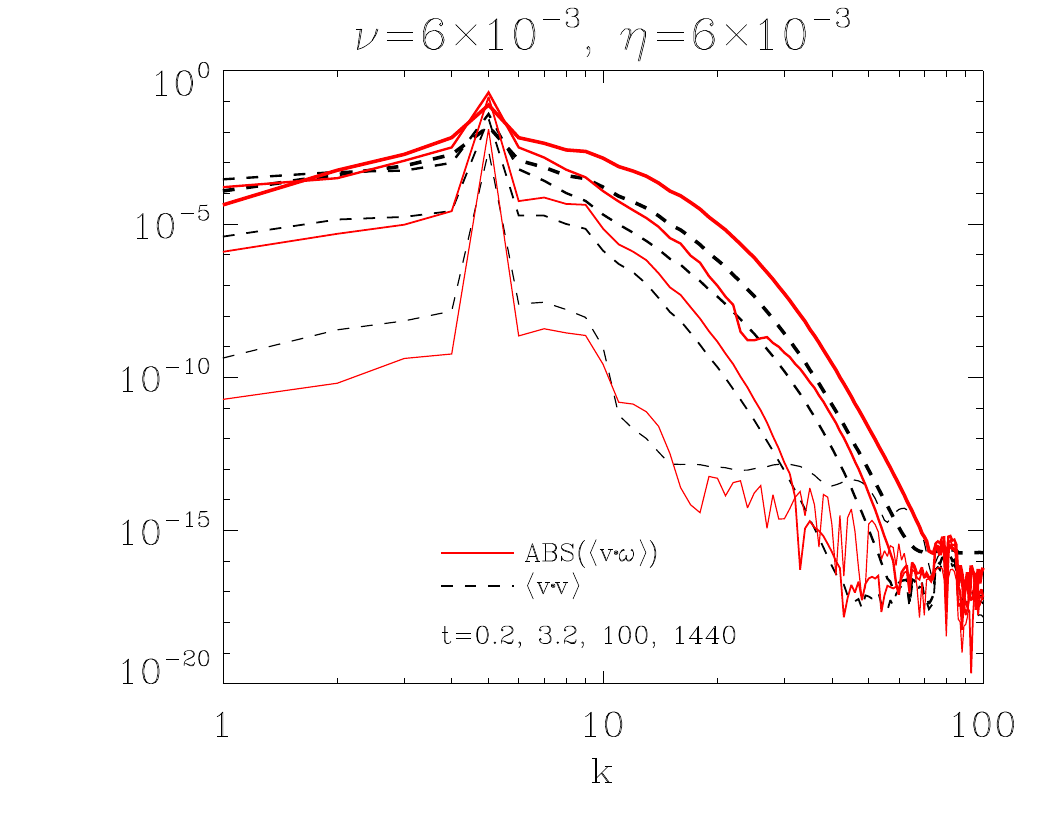}
     \label{f2a}
   }\hspace{-13 mm}
   \subfigure[$|H_M(k)|$, $2E_M(k)$]{
   \includegraphics[width=9.2 cm]{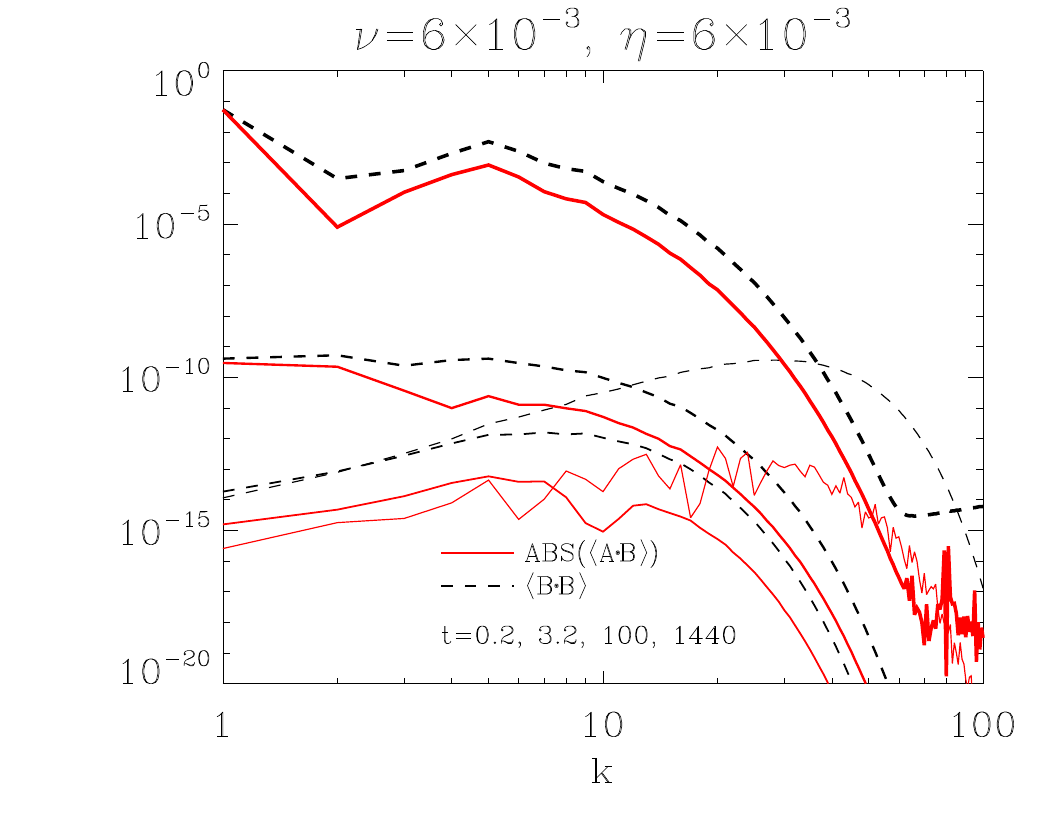}
     \label{f2b}
   }\hspace{-13 mm}
   \subfigure[$K^2H_M(k)$, $2kE_M(k)$]{
     \includegraphics[width=9.2 cm]{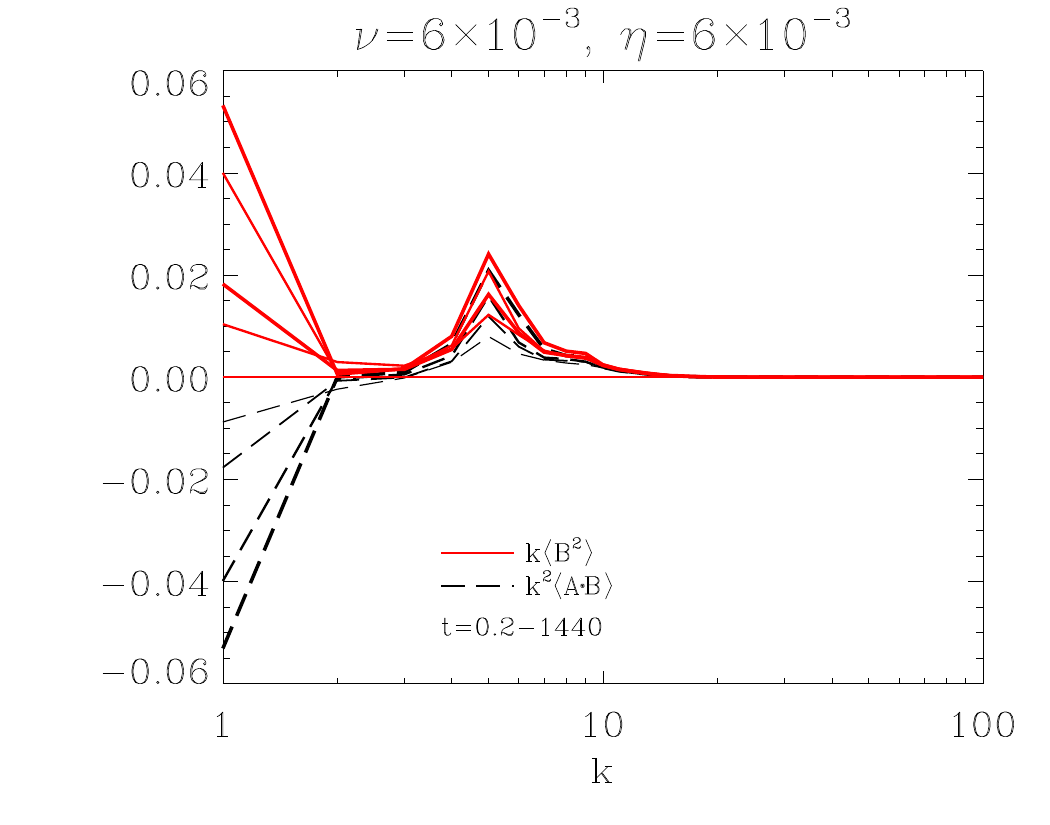}
     \label{f2c}
   }
}
\caption{(a) $\alpha$ \& $\beta$ using Eq.~(\ref{alphaSolution3}), (\ref{betaSolution3}). Magnetic energy and kinetic energy in the large scale regime were multiplied by 10 for clear comparison. (b) Absolute value of $H_V$ is shown because the partial fluctuations in the small scale regime (logarithmic scale). It should be noted that $H_V$ and $E_V$ do not undergo inverse cascaded. Rather, as the figure demonstrates, $H_V$ is more readily forward cascaded. (c) $\overline{E}_M$ $>$ $E_M$ at the forcing scale. The absolute value $|\overline{E}_M|$ was used because its sign is negative, indicating the conservation of magnetic helicity throughout the system. (d) Due to the relatively small values in the small-scale regimes, current helicity $\langle \mathbf{J}\cdot \mathbf{B}\rangle$, along with $2kE_M$, is illustrated on a linear scale.}
\end{figure*}

\begin{figure*}
    {
   \subfigure[$E_M$ \& $E_V$]{
     \includegraphics[width=9.2 cm]{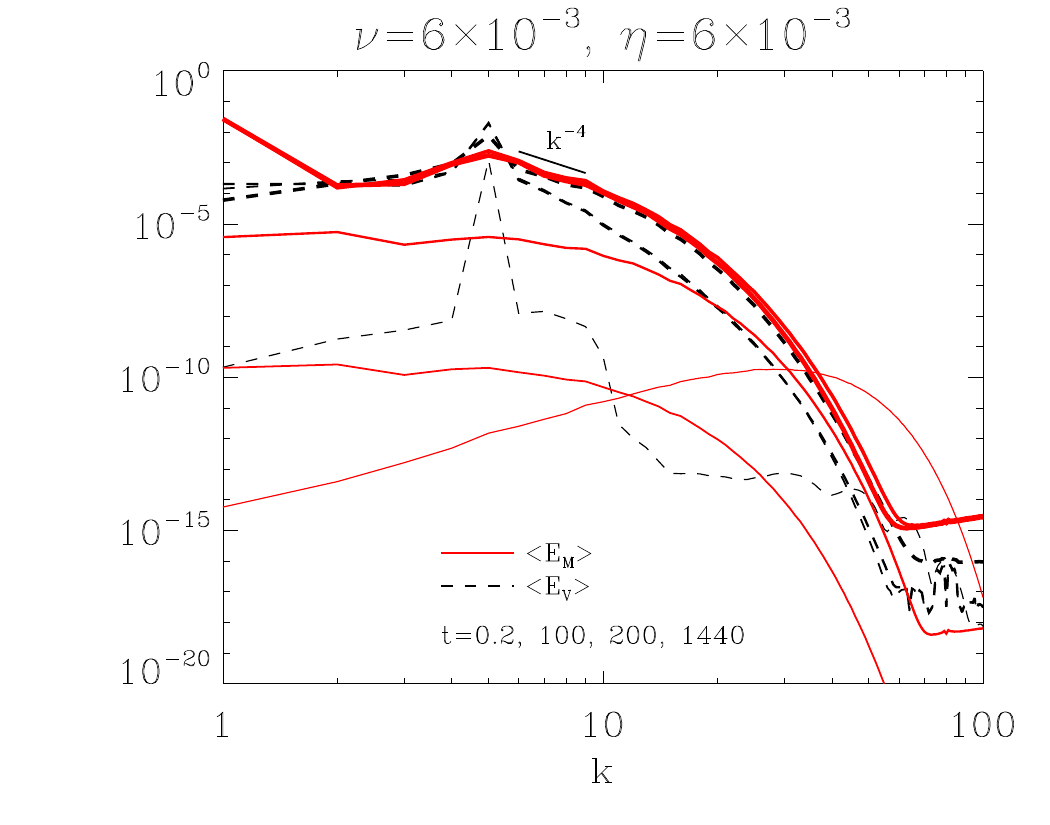}
     \label{f2}
    }\hspace{-13 mm}
   \subfigure[$f_{hk}=\frac{H_V}{2kE_V}$, $f_{hm}=\frac{kH_M}{2E_M}$ at $k=1,\, 5,\, 8$]{
   \includegraphics[width=9.2 cm]{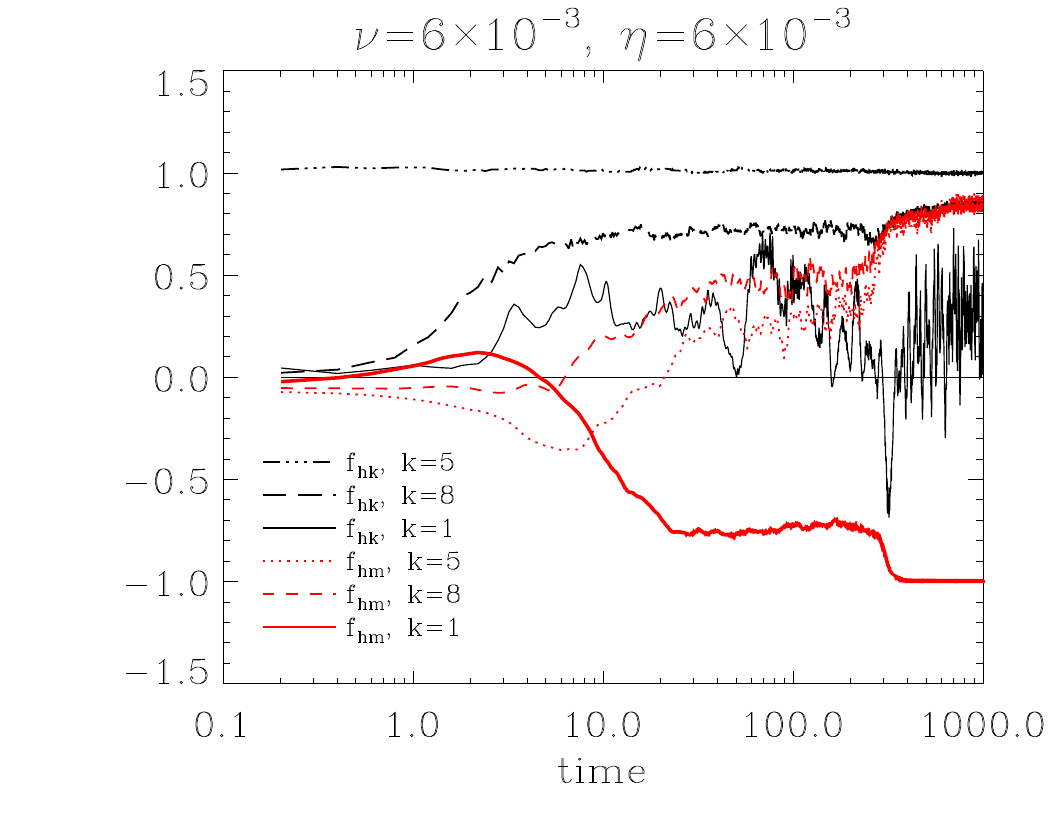}
     \label{f3}
   }\hspace{-13 mm}
   \subfigure[The profiles of $\alpha$ from Eq.~(\ref{Deriv_of_Mean_alpha_beta_MFT_B}) and Eq.~(\ref{alphaSolution3})]{
   \includegraphics[width=9.2 cm]{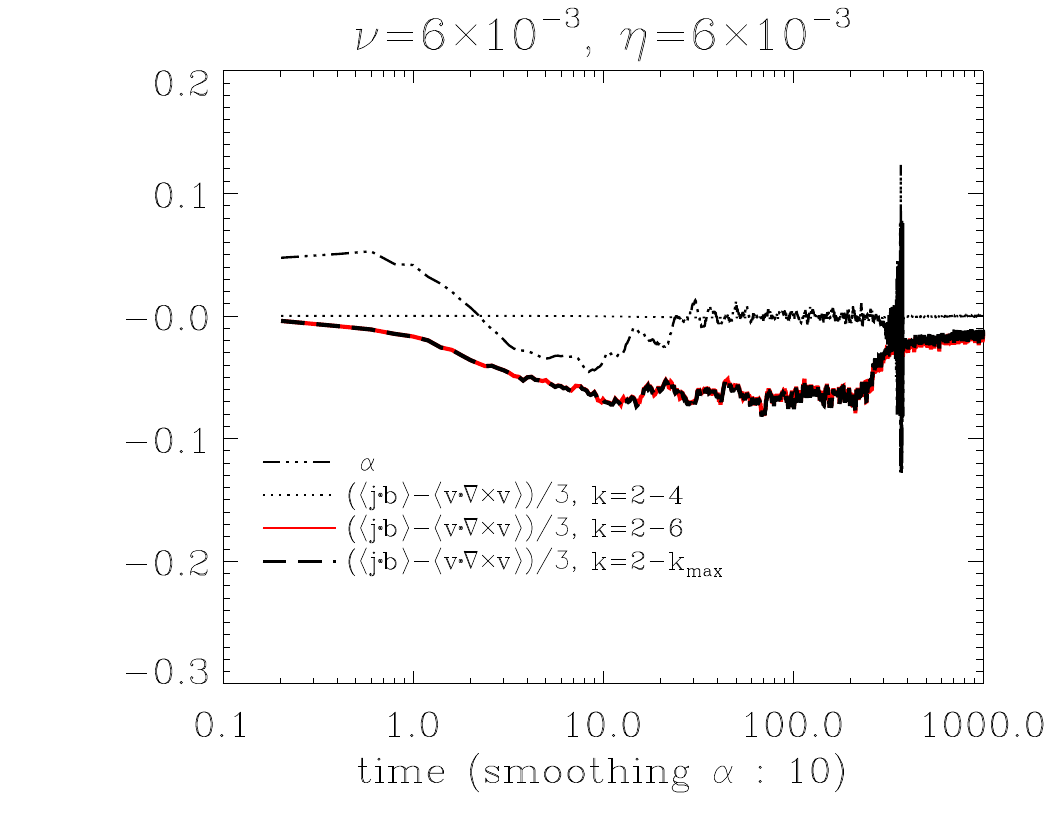}
     \label{f4}
   }\hspace{-13 mm}
   \subfigure[The profiles of $\beta$ from Eq.~(\ref{Deriv_of_Mean_alpha_beta_MFT_B}), (\ref{betaSolution3}), (\ref{general_beta_derivation8})]{
     \includegraphics[width=9.2 cm]{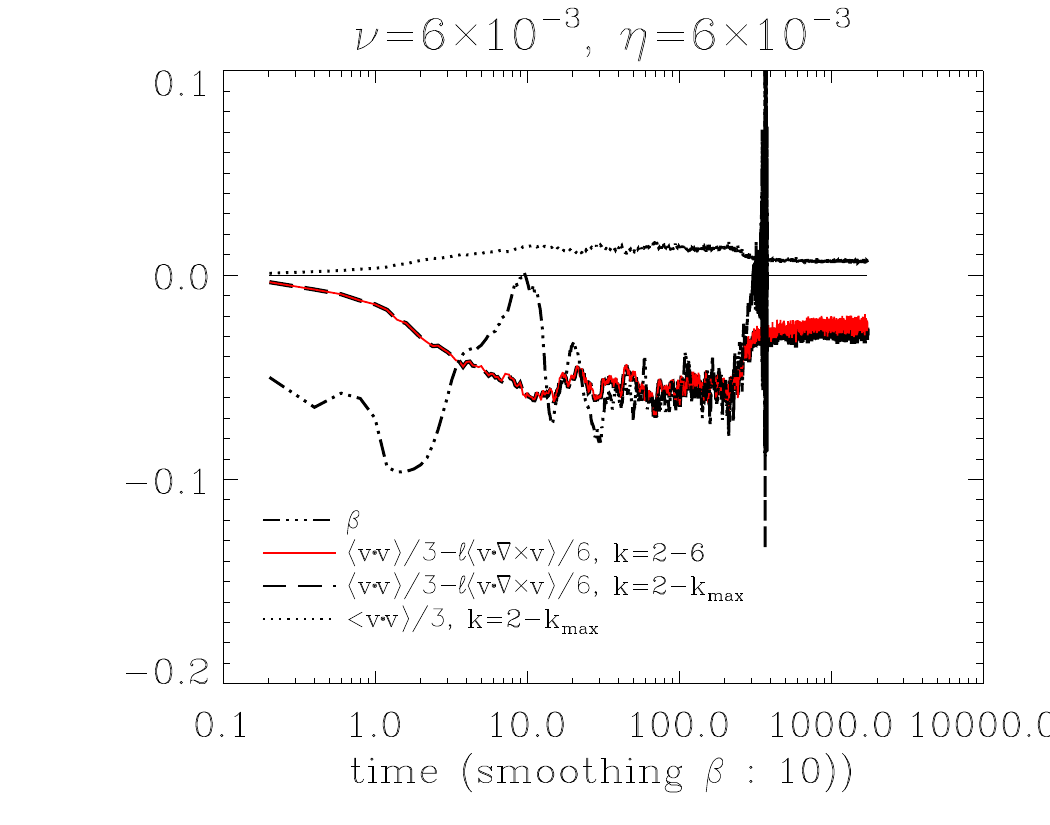}
     \label{f5}
   }
}
\caption{(a) $E_M$ ($k=1$) $ > E_V$ ($k = 5$), implying an inverse cascade of magnetic energy.
(b) $f_{hk}$ represents the kinetic helicity ratio, and $f_{hm}$ represents the magnetic helicity ratio. Also, note that $f_{hm}$ at $k = 1$ is negative. (c) $\alpha$ from Eq.~(\ref{alphaSolution3}) converges to 0 faster than that of MFT, implying the constraint on the effect of induced current on the amplification of magnetic energy $\partial \overline{B}/\partial t\sim \alpha \overline{J}$. (d) Note that without kinetic helicity, $\beta$ from Eq.~(\ref{general_beta_derivation8}) (3 dots-dashed line) becomes the same as $\beta_{MFT}$ from Eq.~(\ref{Deriv_of_Mean_alpha_beta_MFT_B}) (dotted)}
\end{figure*}

\begin{figure*}
    {
   \subfigure[Curl of EMF]{
     \includegraphics[width=8.55 cm]{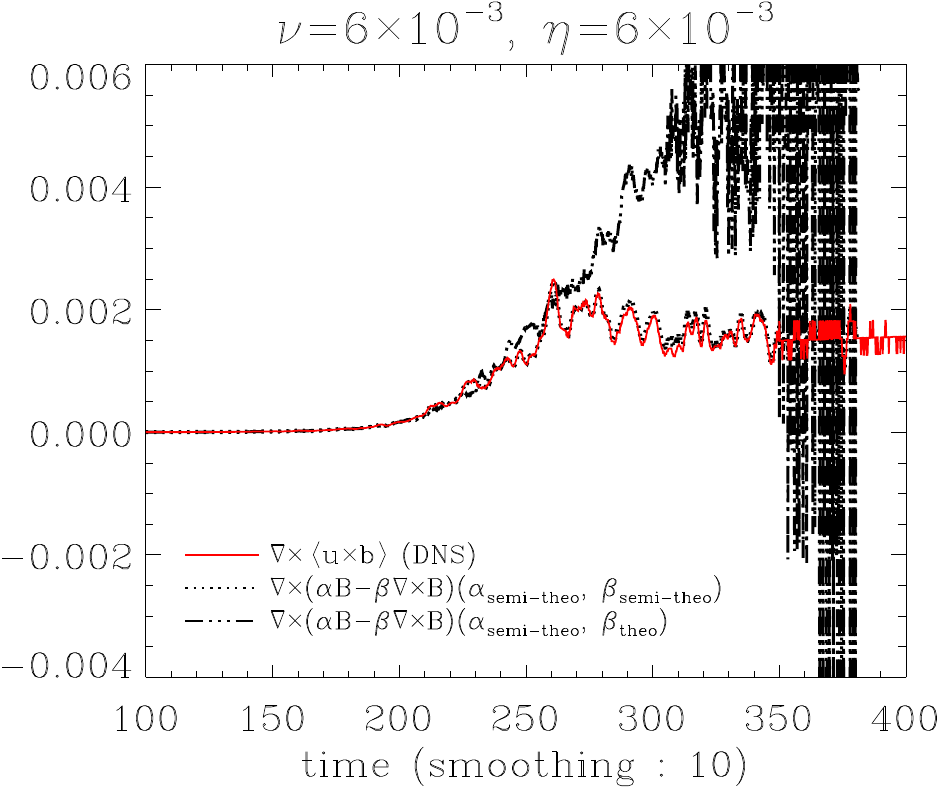}
     \label{f6}
    }\hspace{-0 mm}
   \subfigure[$\frac{\partial \overline{B}}{\partial t}=-\alpha k\overline{B} - (\beta+\eta)k^2\overline{B}$ ($k=1$, $\nabla \times \overline{B}=-\overline{B}$)]{
   \includegraphics[width=8.05 cm]{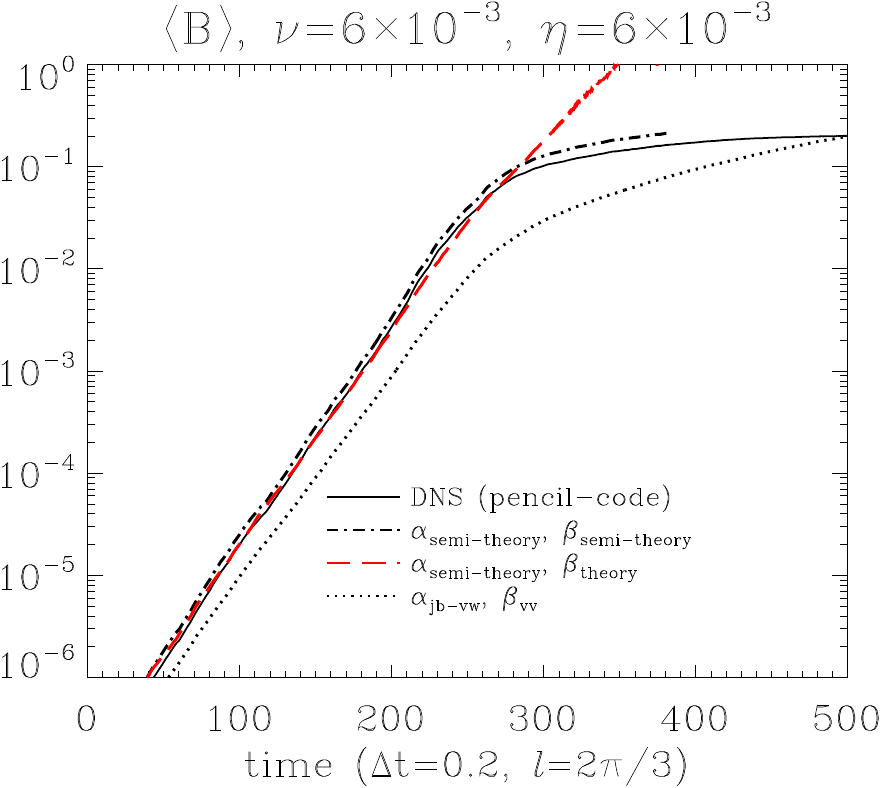}
     \label{f7}
   }
   }
\caption{(a) $\frac{\partial \overline{B}}{\partial t} - \eta \nabla^2 \overline{B}$ was illustrated for $\nabla \times \langle \mathbf{u} \times \mathbf{b} \rangle$ (DNS). Other profiles were illustrated using Eqs.~(\ref{alphaSolution3}), (\ref{betaSolution3}), and (\ref{general_beta_derivation8}). The evolving $\overline{B}$ was reproduced using $\alpha$ and $\beta$ along with the IDL script in the appendix.}
\end{figure*}

\begin{figure*}
    {
   \subfigure[$2\overline{E}$ \& $|\overline{H}_M|$]{
     \includegraphics[width=8.3 cm]{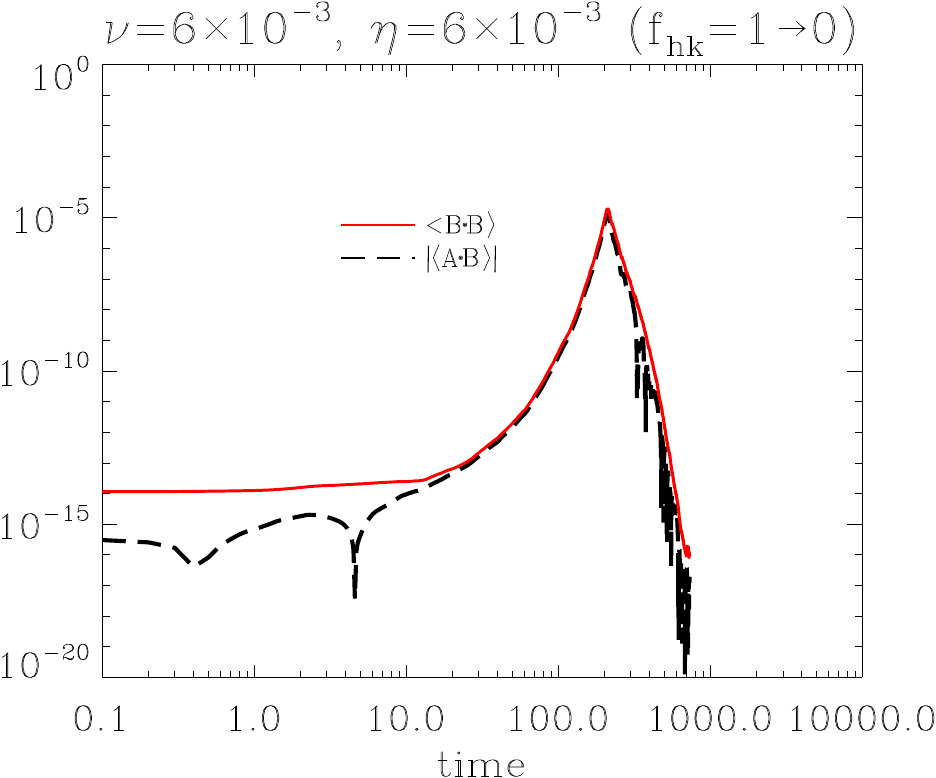}
     \label{f8}   
    }\hspace{-1 mm}
   \subfigure[$f_{hk}$, $f_{hm}$ at $k=1,\, 5,\, 8$]{
   \includegraphics[width=8.0 cm]{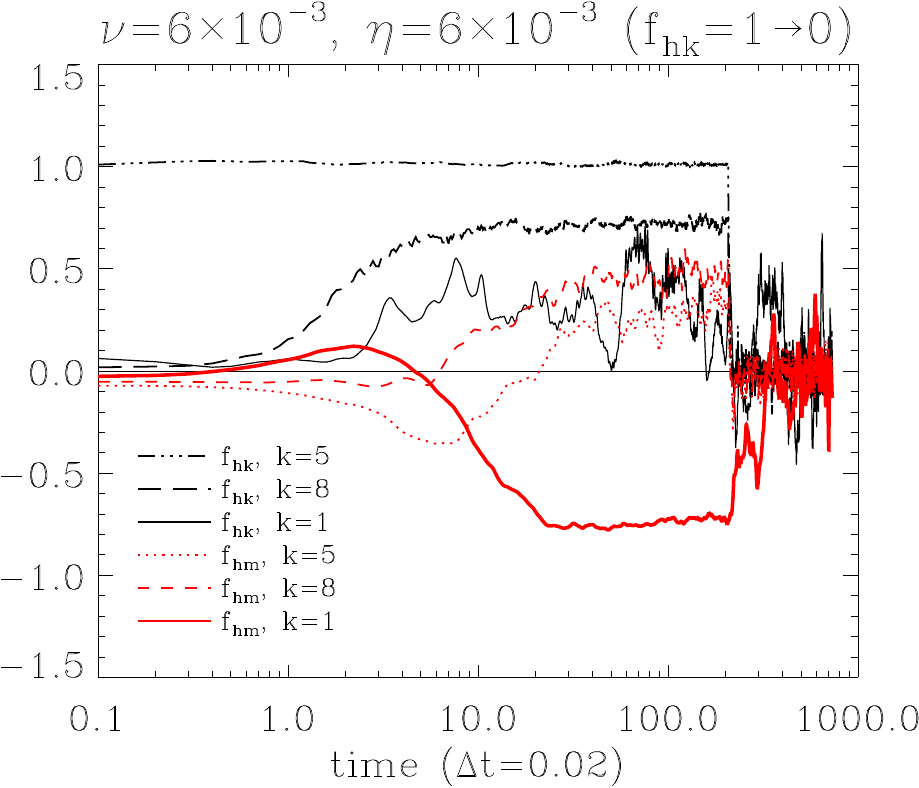}
     \label{f9}  
   }\hspace{-1 mm}
   \subfigure[$\alpha$ vs $\alpha_{MFT}$]{
   \includegraphics[width=8.0 cm]{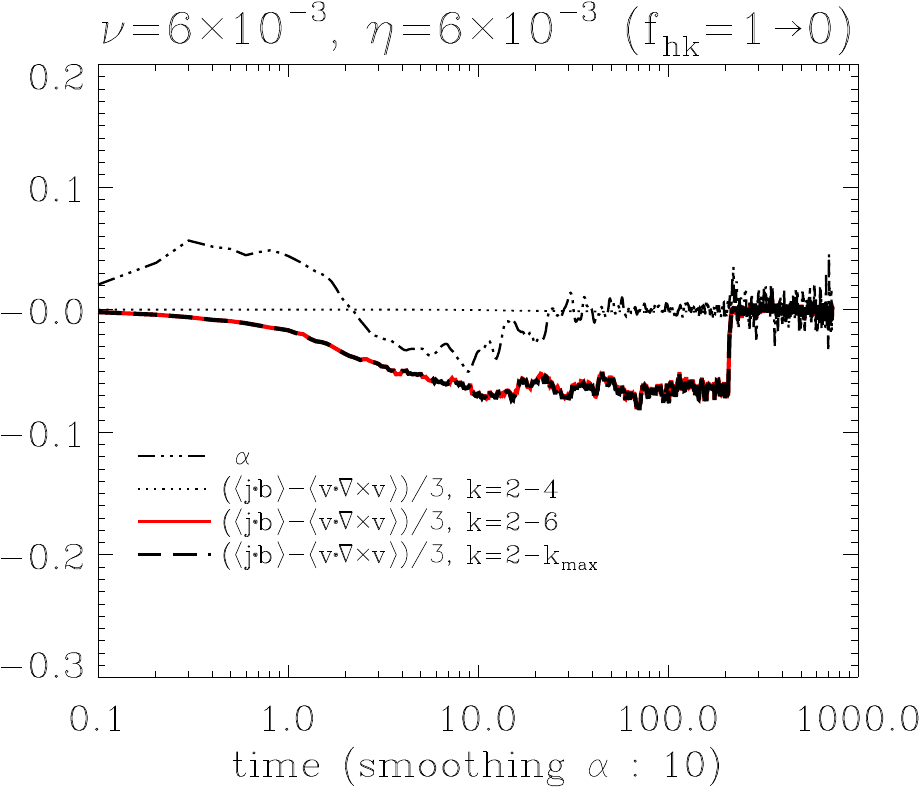}
     \label{f10} 
   }\hspace{-1 mm}
   \subfigure[$\beta$, $\beta_{theo}$, $\beta_{MFT}$]{
     \includegraphics[width=8.1 cm]{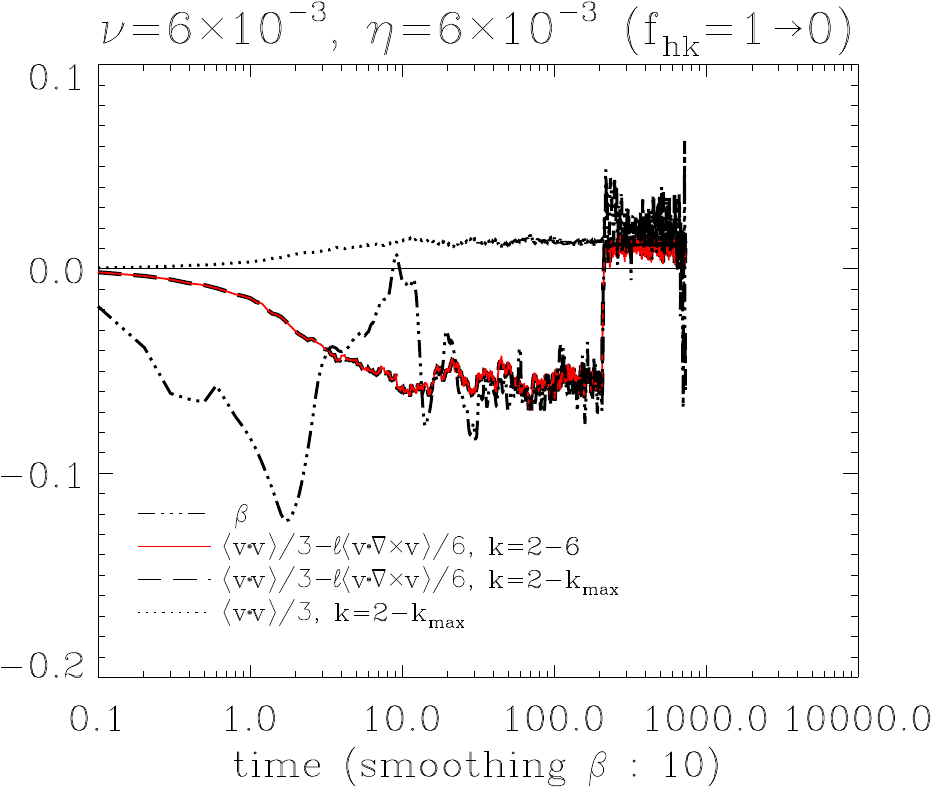} 
     \label{f11}
   }
}
\caption{The basic conditions are the same as those of Figs.~2 and 3, except for the change in helicity in the applied forcing energy. At $t = 210$, the helicity ratio in the forcing function, $f_k$, changed from 1 to 0, while maintaining the same energy.
(a) The large-scale magnetic field begins to drop at $t = 210$, implying no inverse cascade of $E_M$.
(b) All $f_{hk}$ and $f_{hm}$ converge to zero.
(c) $\alpha_{\text{MFT}}$ from Eq.~(\ref{Deriv_of_Mean_alpha_beta_MFT_B}) and $\alpha$ from Eq.~(\ref{alphaSolution3}) converge to zero.
(d) When $f_{hk} = 0$, all $\beta$s from Eqs.~(\ref{Deriv_of_Mean_alpha_beta_MFT_B}), (\ref{betaSolution3}), and (\ref{general_beta_derivation8}) suddenly grow positive and become coincident, implying the role of kinetic helicity in $\beta$.}
\end{figure*}

\begin{figure*}
    {
   \subfigure[$\nabla\times \langle \mathbf{u}\times \mathbf{b} \rangle=\frac{\partial \overline{B}}{\partial t}-\eta\nabla^2\overline{B}$]{
     \includegraphics[width=8.55 cm]{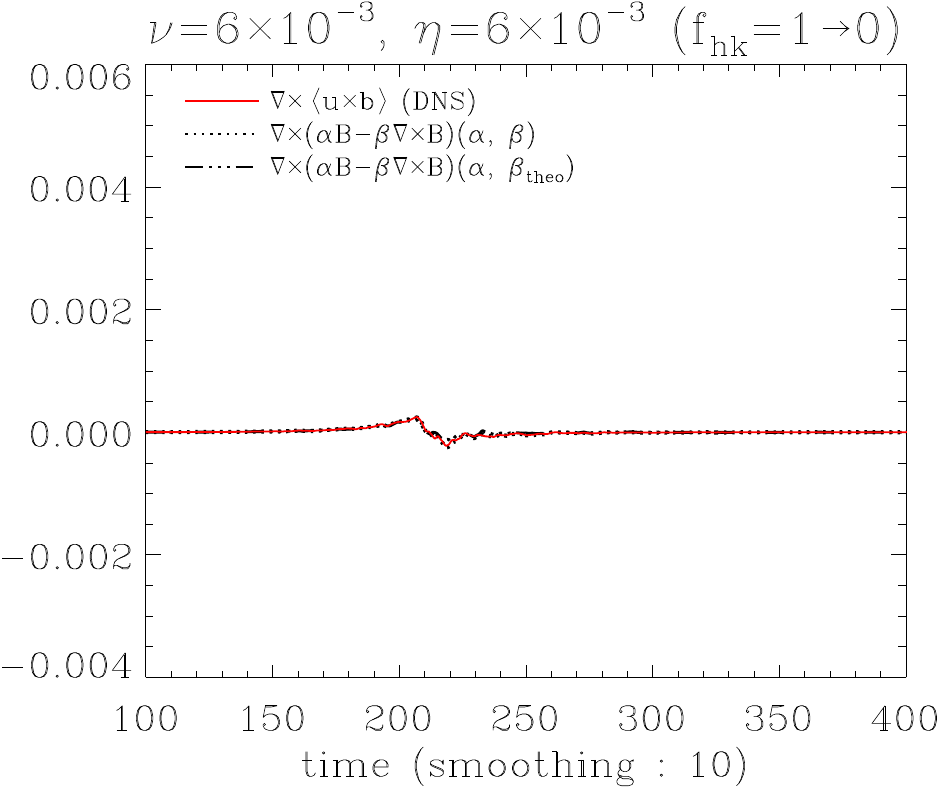}  
     \label{f12}
    }\hspace{-0 mm}
   \subfigure[$\frac{\partial \overline{B}}{\partial t}=-\alpha k\overline{B} - (\beta+\eta)k^2\overline{B}$ ($k=1$, $\nabla \times \overline{B}=-\overline{B}$)]{
   \includegraphics[width=8.25 cm]{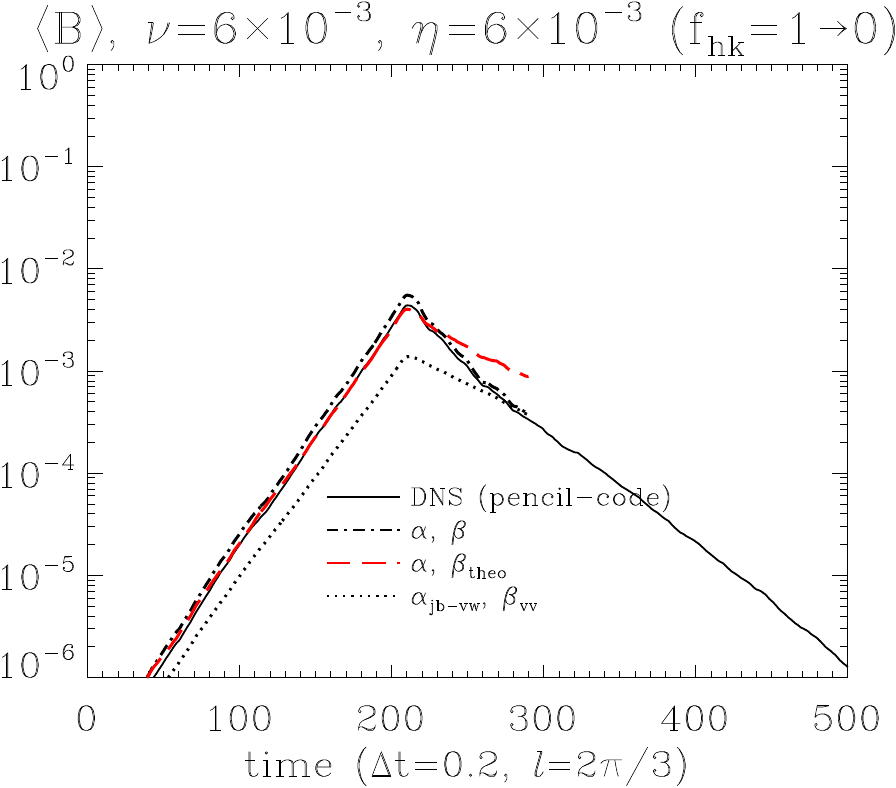}    
     \label{f13}
   }
   }
\caption{(a) Curl of EMFs corresponding to Fig.~\ref{f6}. Note that $\nabla \times \langle \mathbf{u} \times \mathbf{b} \rangle$, which amplifies the large-scale magnetic field, becomes zero in the absence of helicity. (b) The evolution of large-scale magnetic field with and without helicity, in comparison to Fig.~\ref{f7}. These two plots clearly demonstrate the role of kinetic helicity in $\beta$.}
\end{figure*}

\section{Numerical Method}
We employed the $\mathrm{PENCIL\,\,CODE}$ \citep{2001ApJ...550..824B} for our numerical simulations. The computational domain is a periodic cube with dimensions of $(2\pi)^3$, discretized into a mesh of $400^3$ grid points. The code is used to solve the set of magnetohydrodynamic (MHD) equations, which are given as follows:
\begin{eqnarray}
\frac{D \rho}{Dt}&=&-\rho {\bf \nabla} \cdot {\bf U}\label{continuity_equation_Code}\\
\frac{D {\bf U}}{Dt}&=&-c_s^2{\bf \nabla} \mathrm{ln}\, \rho + \frac{{\bf J}{\bf \times} {\bf B}}{\rho}+\nu\big({\bf \nabla}^2 {\bf U}+\frac{1}{3}{\bf \nabla} {\bf \nabla} \cdot {\bf U}\big)+{\bf f}_{kin}\label{momentum_equation_Code}\\
\frac{\partial {\bf A}}{\partial t}&=&{\bf U}{\bf \times} {\bf B} +\eta\,{\bf \nabla}^2{\bf A}\label{vector_induction_equation_Code}\\
\Rightarrow \frac{\partial {\bf B}}{\partial t}&=&\nabla \times ({\bf U}{\bf \times} {\bf B}) +\eta\,{\bf \nabla}^2{\bf B}\label{magnetic_induction_equation_Code}\\
\end{eqnarray}
The symbols used in the equations are defined as follows: `$\rho$' denotes the `density', `$\bf U$' represents the `velocity', `$\bf B$' corresponds to the `magnetic field', `$\bf A$' is the `vector potential', and `${\bf J}$' stands for the `current density'. The symbol `$D/Dt$' refers to the `advective derivative', defined as $\partial / \partial t + {\bf U} \cdot {\bf \nabla}$. The term `$\eta$' (=$c^2/4\pi \sigma$) represents the molecular magnetic diffusivity, where $c$ is the speed of light and $\sigma$ is the conductivity.\\

These equations are presented in dimensionless form. The velocity field and magnetic field are expressed in units of the sound speed `$c_s$' and $(\rho_0\,\mu_0)^{1/2}c_s$, respectively. This scaling is derived from the relations $E_M \sim B^2/\mu_0$ and $E_V \sim \rho_0 U^2$, where `$\mu_0$' and `$\rho_0$' represent the magnetic permeability in free space and the initial density, respectively. It is important to note that the plasma system is weakly compressible, implying `$\rho \sim \rho_0$'. The term `${\bf f}_{kin}(x,t)$' denotes a forcing function, defined as $N\,{\bf f}(t)\,\exp\,[i\,{\bf k}_f(t)\cdot {\bf x} + i\phi(t)]$, where $N$ is a normalization factor, ${\bf f}$ is the forcing magnitude, and ${\bf k}_f(t)$ is the forcing wave number. The code randomly selects one of 20 vectors from the ${\bf k}_f$ vector set at each time step. For simplicity, `$c_s$', `$\mu_0$', and `$\rho_0$' are set to `1', rendering the equations dimensionless.\\

The forcing function ${\bf f}(t)$ is defined as $f_0\mathbf{f}_k(t)$:
\begin{eqnarray}
{\bf f}_k(t)=\frac{i\mathbf{k}(t)\times (\mathbf{k}(t)\times \mathbf{\hat{e}})-\lambda |{\bf k}(t)|(\mathbf{k}(t)\times \mathbf{\hat{e}})}{k(t)^2\sqrt{1+\lambda^2}\sqrt{1-(\mathbf{k}(t)\cdot \mathbf{e})^2/k(t)^2}}.
\label{forcing amplitude fk}
\end{eqnarray}
`$k$' is a wavenumber defined as $2\pi/l$ ($l$: scale size). $k=1$ indicates the large scale regime, and $k>2$ means the wavenumber in the small (turbulent) scale regime. $\lambda=\pm 1$ generates fully right (left) handed helical field $\nabla\times {\bf f}_k\rightarrow i\mathbf{k}\times \mathbf{f}_k\rightarrow \pm k\mathbf{f}_k$. And, `$\mathbf{\hat{e}}$' is an arbitrary unit vector. We gave fully helical kinetic energy ($\lambda=1$) at $\langle k\rangle_{ave}\equiv k_f \sim 5$. This forcing function was located at Eq.~(\ref{momentum_equation_Code}) with $f_0=0.07$ for the helical kinetic forcing dynamo (HKFD). But, $\lambda=0$ yields a nonhelical forcing source. Note that Reynolds rule is not applied to this energy source: $\langle f\rangle \neq 0$.\\

The MHD equation set primarily consists of differential equations, each requiring initial values for their solution. Notably, an initial seed magnetic field of $B_0 \sim 10^{-4}$ was introduced into the system. However, the influence of this seed field diminishes rapidly due to the presence of the forcing function and the lack of memory in the turbulent flow. As we will observe, the small-scale magnetic energy is initially stronger during the very early stages, before decreasing in the subsequent simulation steps.

\section{Numerical Result}
We have driven the plasma system with parameters \(\nu = \eta = 0.006\) using helical kinetic energy input. The helicity ratio is \(f_h = 1\) (fully helical), and the energy is injected at a forcing scale eddy with \(k = 5\). Since our goal was to determine \(\alpha\) and \(\beta\), we selected the most common conditions to minimize unnecessary complexities, such as those caused by an unbalanced dissipation scale with large or small magnetic Prandtl number $Pr_M=\nu/\eta$. In this section, we briefly introduce the numerical results shown in Figs.~1--3, and we will discuss their physical implications using theoretical methods in the next section.\\

Fig.~\ref{f0E} shows the energy evolutions in MHD turbulence dynamo. The externally given kinetic energy drives the plasma at approximately \( U_{\text{rms}} \), and this kinetic motion is converted into magnetic energy \( B_{\text{rms}} \). The initial magnetic field is weak but eventually surpasses the kinetic energy and becomes saturated. The conversion and amplification processes are fundamentally nonlinear and occur through the electromotive force \( \langle \mathbf{U} \times \mathbf{B} \rangle \). Howeve, as Fig.~\ref{f0k} indicates, most energy is located at the forcing scale.\\

Fig.~\ref{f1} shows the large-scale kinetic energy (\(\overline{E}_V\), black dot-dashed line) and magnetic energy (\(\overline{E}_M\), red solid line), both of which are multiplied by 10 for clarity. As expected, \(\overline{E}_M\) rises at \(t \sim 100{-}150\) before reaching saturation, while \(\overline{E}_V\) remains negligibly small. We also include the evolving profiles of \(\alpha\) (dotted line) and \(\beta\) (dashed line) coefficients, obtained from the large-scale magnetic energy and magnetic helicity (\(\overline{H}_M\)). The \(\alpha\) coefficient oscillates, while \(\beta\) remains slightly negative. Both coefficients converge to zero as \(\overline{E}_M\) saturates. Near saturation, \(\alpha\) and \(\beta\) oscillate rapidly in this nonlinear stage, so we applied a smoothing technique of IDL, averaging over approximately 10 points.\\

Fig.~\ref{f2a} presents the spectra of kinetic energy \(E_V\) (black dashed line) and kinetic helicity \(H_V\) (red solid line) in Fourier space at \(t = 0.2\) (lowest), 3.2, 100, and 1440 (highest). The forcing scale is \(k = 5\), and the smallest dissipation scale is \(k_{\text{max}} = 200\). Since the profiles for \(k > 100\) are just extensions of flat noise lines, we cut the spectrum at \(k \sim 100\) for clarity. The plot indicates that kinetic energy injected at the forcing scale cascades toward smaller scales (larger \(k\)). Unlike kinetic energy, kinetic helicity has polarity. In some scale regimes, \(H_V\) is negative, so we plotted its absolute value.\\

Fig.~\ref{f2b} shows the spectra of magnetic energy $2(E_M$ (black dashed line) and magnetic helicity \(H_M\) (red solid line). Similar to Fig.~1(b), we used the absolute value for magnetic helicity, as it also has polarity. Notably, \(\overline{H}_M\) at large scales (\(k = 1\)) is negative, while \(H_M\) at other scales is positive, indicating the conservation of magnetic helicity. The strongest \(\overline{E}_M\), which indicates an inverse cascade, is accompanied by this oppositely polarized \(\overline{H}_M\), characterizing the helical kinetic forcing dynamo. The temporally evolving energy profile at \(k = 1\) is also shown in Fig.~1(a).\\

Fig.~\ref{f2c} includes \( k\langle B^2\rangle \) (red solid) and \( \langle \mathbf{J}\cdot \mathbf{B}\rangle \) (black dashed). Since magnetic fields in the small-scale regime are much weaker compared to those in the large scale, we have used the amplified magnetic helicity and magnetic energy with the wavenumber \( k \). Current helicity \( \langle \mathbf{J}\cdot \mathbf{B}\rangle = k^2\langle \mathbf{A}\cdot \mathbf{B}\rangle \) exhibits similar characteristics to magnetic helicity \( \langle \mathbf{A}\cdot \mathbf{B}\rangle \), apart from the magnitude difference due to the \( k^2 \) factor. This plot illustrates that while the magnetic helicity at the forcing scale is positive, the large-scale magnetic helicity grows to be negative. This demonstrates the conservation of magnetic helicity and the relationship between helicity and energy.\\

Fig.~\ref{f2} displays \(E_V\) (black dashed line) and \(E_M\) (red solid line) together. It is evident that the kinetic energy injected at \(k = 5\) is converted into magnetic energy and undergoes an inverse cascade to larger scales. Due to the complex energy transfer, Kolmogorov's \(k^{-5/3}\) scaling is not observed; instead, the spectrum is much steeper, approximately \(k^{-4}\). These processes are inherently nonlinear, making them challenging to understand. However, they may be linearized using the \(\alpha\) and \(\beta\) parameters to make the dynamics more intuitively understandable.\\


Fig.~\ref{f3} shows the helicity ratios. The kinetic helicity ratio, \(f_{hk}\), was calculated using \(\langle \mathbf{U} \cdot \boldsymbol{\omega} \rangle / k \langle U^2 \rangle\), where \(\boldsymbol{\omega}\) is the vorticity \(\nabla \times \mathbf{U}\). The magnetic helicity ratio, \(f_{hm}\), was calculated using \(k\langle \mathbf{A} \cdot \mathbf{B} \rangle / \langle B^2 \rangle\). The helicity ratios were computed for \(k = 1\)(large scale), \(k = 5\)(forcing small scale), and \(k = 8\)(small scale). Since we applied fully helical kinetic energy, \(f_{hk}\) at \(k = 5\) remains 1. The other kinetic helicity ratios behave similarly to that at the forcing scale, except at the large scale. For magnetic helicity, the polarity depends on the wavenumber. \(f_{hm}\) at \(k = 1\) (large scale) saturates at `-1', while \(f_{hm}\) at smaller scales, \(k = 5\) and \(k = 8\), becomes positive. These opposite signs clearly demonstrate the conservation of magnetic helicity. Of course, since we externally impose helical energy, the total magnetic helicity is not exactly zero. However, the tendency toward conservation is maintained. And, it should be noted that a helicity ratio less than 1 indicates the production of nonhelical components, which is a natural occurrence in turbulence.\\


Fig.~\ref{f4} shows and compares the profiles of \(\alpha\) and \(\frac{1}{3} \left(\langle \mathbf{j} \cdot \mathbf{b} \rangle - \langle \mathbf{u} \cdot \boldsymbol{\omega} \rangle \right)\). The \(\alpha\) profile was calculated using \(\overline{E}_M\) and \(\overline{H}_M\). The \(\alpha\) profile from the large-scale magnetic data oscillates and converges to zero early. In contrast, the conventional integrand constituting \(\alpha\)\footnote{For simplicity, we assumed a correction time of 1, such that \(\int^{\tau} f \, dt \rightarrow f\tau \rightarrow f\).} remains negative for a longer period and saturates at a decreased negative value as \(\overline{B}\) saturates. Since the system was driven with helical kinetic energy, kinetic helicity is naturally larger than current helicity. We tested the profiles for \(k=2-4\), \(k=2-6\), and \(k=2-k_{\text{max}}\). The forcing scale regime, including \(k=5\), primarily determines the profile.\\

Fig.~\ref{f5} shows the profiles of \(\beta\) derived from large-scale magnetic data (\(\overline{E}_M\) and \(\overline{H}_M\)) and from small-scale kinetic data \(\frac{1}{3} \langle u^2 \rangle - \frac{l}{6} \langle \mathbf{u} \cdot \boldsymbol{\omega} \rangle\). For simplicity, we also set the correlation time to 1. Compared to the conventional \(\beta\), the latter includes the effect of kinetic helicity modified by the correlation length \(l/6\). Since the exact method for determining \(l\) is not yet known, we used the inverse of the wavenumber \(2\pi/k\). At present, the smallest wavenumber in the small-scale regime, \(k = 2\), provides the best fit to the \(\beta\) profiles. When larger wavenumbers are used, the profiles tend to increase and approach zero. Since a given eddy cannot simultaneously possess both toroidal and poloidal components, it can be inferred that there exists a correlation length between the toroidal and poloidal components that constitute the kinetic helicity. Although we have used a trial-and-error method to find \(l\), a more precise method and physical analysis are necessary.\\

Fig.~\ref{f6} shows a comparison between the electromotive force (EMF) \(\langle \mathbf{u} \times \mathbf{b} \rangle\) and the linearized EMF, given by \(\alpha \mathbf{B} - \beta \nabla \times \mathbf{B}\). The exact profiles of \(\langle \mathbf{u} \times \mathbf{b} \rangle\) are not available at present due to the uncertainty in the exact range of the small-scale regime. Instead, we used \(\partial \overline{\mathbf{B}} / \partial t - \eta \nabla^2 \overline{\mathbf{B}}\) (red solid line) as a substitute for \(\langle \mathbf{u} \times \mathbf{b} \rangle\) in the small scale regime. This serves as a reasonable approximation for the EMF at small scales. For \(\alpha\), we used the large-scale magnetic data \(\overline{H}_M\) and \(\overline{E}_M\) for this semi \(\alpha\) instead of the conventional approach. And, for semi \(\beta\)(dotted line), one version is derived from the large-scale magnetic data like $\alpha$, while theo $\beta$ (dot-dashed line) comes from a theoretical approach that includes kinetic helicity and kinetic energy. The EMF grows and saturates around \(0.002\), which is compensated by \(\eta \nabla^2 \overline{\mathbf{B}}\) to ensure that \(\partial \overline{\mathbf{B}} / \partial t\) approaches zero. Initially, the linearized EMF with large-scale magnetic data (semi \(\alpha\) and semi \(\beta\) as well as theo \(\beta\)) closely matches \(\langle \mathbf{u} \times \mathbf{b} \rangle\). {However, as the nonlinear stage progresses or magnetic effects grow, the EMF with $(\alpha_{\text{semi-theo}},\, \beta_{\text{theo}})$ becomes larger than that of the other set. This indicates that $\beta_{\text{theo}}$ is not sufficiently quenched. In contrast, the EMF with $(\alpha_{\text{semi-theo}},\, \beta_{\text{semi-theo}})$ aligns precisely with $\nabla \times \langle \mathbf{u} \times \mathbf{b} \rangle$.} Due to the significant numerical noise, a smoothing function in IDL (averaging over 10 neighboring points) was applied. In contrast, theo-\(\beta\) derived from the small-scale kinetic data performs better during this nonlinear stage.\\

Fig.~\ref{f7} includes the reproduced the large-scale magnetic field \(\overline{\mathbf{B}}\) using two approaches: (semi-\(\alpha\), semi-\(\beta\), black dot-dashed line) and (semi-\(\alpha\), theo-\(\beta\), red dashed line). These are compared with the numerically calculated \(\overline{\mathbf{B}}\) from the code (black solid line). The reproduced fields match quite well with the DNS data, but in the nonlinear stage, the theoretical \(\beta\) yields better results than the semi-analytic \(\beta\). As noted, during this stage, numerical oscillations caused by the close values of \(\langle\overline{\mathbf{B}}^2\rangle\) and \(\langle\overline{\mathbf{A}} \cdot \overline{\mathbf{B}}\rangle\) increase significantly.\\

{Figs.~5, 6 are designed to examine the dependency of $\alpha$ and $\beta$ on (kinetic) helicity from different perspectives. Fig.~ \ref{f8} illustrates the evolution of the profiles of $2\overline{E}_M$ and $\overline{H}_M$ in a system driven by helical kinetic energy (at $k=5$) for $t<210$, and by nonhelical kinetic energy (also at $k=5$) for $t>210$. Here, helicity is controlled by adjusting $\lambda$ in Eq.~\ref{forcing amplitude fk} from $\lambda=1$ to $0$, while keeping the energy constant. When kinetic helicity vanishes at $t=210$, $\overline{B}$ undergoes a sharp decline as the figure shows. Fig.~\ref{f9} shows the changes in the helicity ratio of kinetic eddies and magnetic eddies across different scales. In the $t<210$ interval, the helical forcing case is similar to Fig.~\ref{f3}; however, after helicity is removed, the helicity ratio converges to zero in many cases. Nonetheless, at small scales, a nonzero region persists for some time. In Fig.~\ref{f10}, the evolution of $\alpha$ and $\alpha_{MFT}$ is presented, where it is notable that $\alpha_{MFT}$ converges to zero almost simultaneously as kinetic helicity disappears. $\alpha$ has already approached zero, so its change is minimal. However, in the case of $\alpha_{MFT}$, as shown in Fig.~\ref{f9}, $\langle {\bf j} \cdot {\bf b} \rangle$ remains nonzero for a time, but $\alpha_{MFT}$ converges to zero more rapidly. This suggests that the defining equation for MFT, Eq.~(\ref{Deriv_of_Mean_alpha_beta_MFT_B}), may be insufficient. Fig.~\ref{f11} compares $\beta$, $\beta_{theo}$, and $\beta_{MFT}$. It is noteworthy that all three measures converge when helicity is absent.\\}

{Fig.~\ref{f12} compares the source of $\overline{B}$, $\nabla \langle {\bf u} \times {\bf b} \rangle$, with the EMF composed of $\alpha$ and $\beta$. Since helicity was removed before the EMF sufficiently increased, the changes are minor, but it is presented here in the same format as Fig.~\ref{f6} for consistency. Interestingly, as helicity supply ceases, the curl of EMF converges to zero, yet slight oscillations remain. Fig.~\ref{f13} reconstructs $\overline{B}$ using $\alpha$ and $\beta$, confirming a consistent result.}

\section{Theoretical approach}
\subsection{Conventional Derivation of $\alpha$ \& $\beta$}
With Reynolds rule $\langle XY \rangle = \overline{X}\overline{Y} + \cancel{\langle\overline{X}y\rangle} + \cancel{\langle x\overline{Y}\rangle}+ \langle xy \rangle$, the large scale magnetic induction equation is represented as
\begin{eqnarray}
\frac{\partial \overline{\bf B}}{\partial t}&=&\nabla \times \langle {\bf u}\times {\bf b}\rangle-\eta \nabla \times \nabla \times \overline{\bf B}.
\label{magnetic_induction_EMF}
\end{eqnarray}
The electromotive force (EMF) $\langle {\bf u}\times {\bf b}\rangle$ is inherently nonlinear, making precise analytic calculations challenging. However, the EMF can be approximately linearized using the parameters $\alpha$, $\beta$, and the large-scale magnetic field $\overline{\bf B}$: $\langle \mathbf{u}\times \mathbf{b}\rangle \sim \alpha \overline{\mathbf{B}} - \beta \nabla \times \overline{\mathbf{B}}$. Consequently, the equation can be rewritten as:
\begin{eqnarray}
\frac{\partial \overline{\bf B}}{\partial t}&\sim& \nabla\times (\alpha\overline{\bf B}-\beta \nabla \times \overline{\bf B)}-\eta \nabla \times \nabla \times \overline{\bf B}\nonumber\\
&\sim&\alpha \overline{\bf J} + (\beta+\eta) \nabla^2 \overline{\bf B}.
\label{magnetic_induction_alpha_beta}
\end{eqnarray}
This form is obtained by differentiating Eq.~(\ref{magnetic_induction_equation_Code}) with respect to time and then recursively applying Eq.~(\ref{momentum_equation_Code}) and Eq.~(\ref{magnetic_induction_equation_Code}). During this process, some higher-order or nonlinear terms may be neglected or not fully calculated. However, this equation retains a complete form with respect to the generation of magnetic fields from an electromagnetic perspective. Eq.~(\ref{magnetic_induction_alpha_beta}) suggests that the magnetic field in the plasma system is induced by the current density $\overline{\bf J}$ through the $\alpha$ effect, in accordance with Ampère's law. While this is fundamentally electrodynamic, it can be related to the static Biot-Savart law. Additionally, the interactions between magnetic fields and numerous charged particles with mass necessitate an additional term, $\nabla^2 \overline{\bf B}$, associated with the $\beta$ effect. This arises from the relation $\nabla \times \overline{\bf J} = \nabla \nabla \cdot \overline{\bf B} - \nabla^2 \overline{\bf B}$, which is mathematically analogous to diffusion. While the concept of fluidic diffusion is applied in MHD, it is desirable to retain its physical significance.\\

\subsubsection{Mean Field Theory}
In Mean Field Theory (MFT), $\bf u$ and $\bf b$ are substituted with $\int \partial {\bf u}/\partial t\, dt$ and $\int \partial {\bf b}/\partial t\, dt$, respectively. The small-scale momentum and magnetic induction equations are then approximated accordingly \citep{1980opp..bookR....K, 2008matu.book.....B}.
\begin{eqnarray}
\frac{\partial {\bf u}}{\partial t}\sim\overline{\bf B}\cdot\nabla{\bf b},\quad \frac{\partial {\bf b}}{\partial t}\sim-{\bf u}\cdot \nabla\overline{\bf B}+\overline{\bf B}\cdot\nabla{\bf u}.
\label{turbulentMHD}
\end{eqnarray}
Then,
\begin{eqnarray}
{\bf u}\times {\bf b}&\sim& \int^t \overline{\bf B}\cdot\nabla {\bf b}\,d\tau\,\times \,{\bf b}+{\bf u}\,\times \int^t (-{\bf u}\cdot\nabla \overline{\bf B}+\overline{\bf B}\cdot\nabla {\bf u})\,d\tau\\
&\sim& \epsilon_{ijk}\int^t \overline{B}_l\nabla_l {b_j}\,d\tau \,{b_k}+\epsilon_{ijk}{u_j}\int^t (-{u_l}\nabla_l \overline{B}_k+\overline{B}_l\nabla_l {u_k})\,d\tau.
\label{Deriv_of_Mean_alpha_beta_MFT_A}
\end{eqnarray}
In a homogeneous and isotropic system, the tensor identities $\langle x_k\partial_ix_j - x_j\partial_ix_k \rangle = \frac{1}{3} \langle \mathbf{x} \cdot \nabla \times \mathbf{x} \rangle$ and $\langle u_ju_l \rangle = \frac{1}{3} \langle u^2 \rangle$ can be applied. Therefore,
\begin{eqnarray}
\langle {\bf u}\times {\bf b} \rangle \sim \underbrace{\bigg(\frac{1}{3}\int^t \big(\langle {\bf j}\cdot {\bf b}\rangle-\langle {\bf u}\cdot \nabla\times {\bf u}\rangle\big)\, d\tau\bigg)}_{\alpha_{MFT}}\overline{\bf B}-\underbrace{\bigg(\frac{1}{3}\int^t \langle u^2\rangle\, d\tau\bigg)}_{\beta_{MFT}}\nabla \times \overline{\bf B}.
\label{Deriv_of_Mean_alpha_beta_MFT_B}
\end{eqnarray}
Rigorously, $\langle x_ix_j\rangle$ indicates $\langle x_i(r,\,t)x_j(r+\delta r,\,t+\delta t)\rangle$. The replacement of $\langle u_ju_l\rangle$ with $1/3\langle u^2\rangle$ is somewhat oversimplified.\\

\subsubsection{Direct Interaction Approach}
In the Direct Interaction Approximation (DIA, \cite{Akira2011}), a second-order statistical relation is employed in place of the vector identity typically used in Mean Field Theory (MFT).
\begin{eqnarray}
\langle X_i({k})X_j({-k})\rangle=\big(\delta_{ij}-\frac{k_ik_j}{k^2} \big)E_X(k)+\frac{i}{2}\frac{k_l}{k^2}\epsilon_{ijl}H_X(k)\label{Statistical_relation}\\
\bigg(\langle X^2\rangle=2\int E_X(k)d{k},\,\,\langle {\bf X}\cdot \nabla\times {\bf X}\rangle=\int H_X(k)d{k}\bigg)\nonumber
\end{eqnarray}
With the Green's function $G$ incorporating higher-order nonlinear terms, EMF is represented by $\alpha$, $\beta$, and $\gamma$ for cross helicity $\langle {\bf u}\cdot {\bf b}\rangle$:
\begin{eqnarray}
\alpha_{DIA}&=&\frac{1}{3}\int d{\bf k}\int^t G\big(\langle {\bf j}\cdot {\bf b} \rangle-\langle {\bf u}\cdot \nabla\times {\bf u}\rangle\big) d\tau, \label{DIA_alpha}\\
\beta_{DIA}&=&\frac{1}{3}\int d{\bf k}\int^t  G\big(\langle u^2\rangle+\langle b^2\rangle\big) d\tau,
\label{DIA_beta}\\
\gamma_{DIA}&=&\frac{1}{3}\int d{\bf k}\int^t G\langle {\bf u}\cdot {\bf b} \rangle\, d\tau.\label{DIA_gamma}
\end{eqnarray}
Compared to Mean Field Theory (MFT), the $\gamma$ effect is included alongside $\alpha$ and $\beta$ within the Green's function $G$. Additionally, $\beta$ incorporates both turbulent kinetic energy and magnetic energy.\\

\subsubsection{Eddy damped quasinormal Markovian theory}
The coefficients $\alpha$ and $\beta$ can be also calculated using EDQNM approach. THe fourth-order moments $\langle x_l x_m x_n x_q \rangle$ appear in the calculations. These higher-order moments are approximated by products of second-order moments, such as $\sum\langle x_lx_m \rangle\langle x_nx_q \rangle$, a process known as quasi-normalization. These second-order moments are then expressed in terms of energy $E$, helicity $H$, and cross helicity $\langle{\bf u}\cdot {\bf b}\rangle$ using the relations from Eq.(\ref{Statistical_relation}) \citep{1967PhFl...10..859K, 1976JFM....77..321P, 1990cp...book.....M}. While the basic concept is straightforward, the detailed calculations are quite much. Here, we simply present the results \citep{1976JFM....77..321P}:\\
\begin{eqnarray}
\alpha_{QN}&=&\frac{2}{3}\int^{t} \Theta_{kpq}(t)\big(\langle {\bf j}\cdot {\bf b} \rangle- \langle {\bf u}\cdot \nabla\times {\bf u}\rangle\big)\,dq,
\label{EDQNM_alpha}\\
\beta_{QN}&=&\frac{2}{3}\int^{t} \Theta_{kpq}(t)\langle u^2 \rangle\,dq.
\label{EDQNM_beta}
\end{eqnarray}
A triad relaxation time, $\Theta_{kpq}$, is defined as $\Theta_{kpq} = \frac{1 - \exp(-\mu_{kpq}t)}{\mu_{kpq}}$, where the eddy damping operator $\mu_{kpq}$ must be determined experimentally. In a stable system, the relaxation time converges to a constant value over time: $\Theta_{kpq} \sim \mu^{-1}_{kpq} \rightarrow \text{const}$. Notably, the coefficients of $\alpha$ and $\beta$ are $2/3$, which is a fundamental result of the quasi-normalization process that approximates fourth-order moments by combinations of second-order moments. Furthermore, when quasi-normalization is applied to nonlinear moments, additional energy and helicity terms are introduced. In principle, this effect is regulated by $\Theta_{kpq}$ and $\mu_{kpq}$.\\

The differences in $\alpha$ and $\beta$ across MFT, DIA, and EDQNM indicate that the linearization of the electromotive force (EMF) depends on the closure theory applied. However, they commonly show similar residual helicity in $\alpha$ and energy in $\beta$. in Fig.~\ref{f4}, we have compared $\alpha$ derived from $\overline{H}_M$ and $\overline{E}_M$ with the residual helicity $\langle \mathbf{j} \cdot \mathbf{b} \rangle - \langle \mathbf{u} \cdot \nabla \times \mathbf{u} \rangle$, assuming a unit correlation time. The comparison shows that the forcing scale $k=5$ substantially determines the profile, and some other effects are missing.\\


\subsection{Derivation of $\alpha$ \& $\beta$ from $\overline{E}_M$ \& $\overline{H}_M$}
In the previous sections, we introduced the calculation of $\alpha$ and $\beta$ coefficients using MFT, DIA, and EDQNM models. While these models help in understanding the qualitative aspects, they make it difficult to determine how the $\alpha$ and $\beta$ coefficients actually evolve. Therefore, we sought a method to determine the profiles of $\alpha$ and $\beta$ using readily measurable data. We have previously worked on this model\citep{2023ApJ...944....2P}, but we believe it is worthwhile to introduce it for application to the DNS data.\\

As discussed, Eq.~(\ref{magnetic_induction_alpha_beta}) is formally complete. It indicates that the magnetic field is induced by current density and diffusion. These two effects encompass the most essential properties in plasmas dominated by statistical electromagnetism. From Eq.(\ref{magnetic_induction_alpha_beta}), we get
\begin{eqnarray}
\overline{\mathbf{A}}\cdot \frac{\partial \overline{\mathbf{B}}}{\partial t}
&=& \alpha \overline{\mathbf{A}}\cdot \nabla\times \overline{\mathbf{B}}+(\beta+\eta) \overline{\mathbf{A}}\cdot \nabla^2\overline{\mathbf{B}}\nonumber\\
&\rightarrow& \alpha \overline{B}^2-(\beta+\eta)\, \overline{\mathbf{A}}\cdot \overline{\mathbf{B}}\\
\overline{\mathbf{B}}\cdot \frac{\partial \overline{\mathbf{A}}}{\partial t}
&=& \alpha \overline{\mathbf{B}}\cdot \nabla\times \overline{\mathbf{A}}+(\beta+\eta) \overline{\mathbf{B}}\cdot \nabla^2\overline{\mathbf{A}}\nonumber\\
&\rightarrow&  \overline{B}^2-(\beta+\eta)\, \overline{\mathbf{A}}\cdot \overline{\mathbf{B}}
\end{eqnarray}
Then, we have
\begin{eqnarray}
\frac{d}{dt}\overline{H}_M&=&4\alpha \overline{E}_M-2(\beta+\eta)\overline{H}_M.\label{Hm1}
\label{Hm1}
\end{eqnarray}
This equation is simplified with \( k = 1 \), which corresponds to the large scale magnetic field $\overline{\mathbf{B}}$, relative to the system\footnote{Note that the large eddy scale for $k=1$ is not absolute but relative. The evolution of the magnetic field under the influence of the \(\alpha\) and \(\beta\) effects is valid only on large scales.}.\\

In the same way, we get the magnetic energy in the large scale:
\begin{eqnarray}
\frac{\partial}{\partial t}\overline{E}_M&=& \alpha {\overline H}_M-2(\beta+\eta) {\overline E}_M.
\label{Em1}
\end{eqnarray}
These coupled differential equations can be easily solved through diagonalization:
\begin{eqnarray}
\left[
\begin{array}{c}
\frac{\partial \overline{H}_M}{\partial t} \\
\frac{\partial \overline{E}_M}{\partial t}
\end{array}
\right]
=\left[
\begin{array}{cc}
-2(\beta+\eta) & 4\alpha\\
     \alpha         &-2(\beta+\eta)
\end{array}
\right]
\left[
\begin{array}{c}
\overline{H}_M \\
\overline{E}_M
\end{array}
\right]
=
\left[
\begin{array}{cc}
\lambda & 0\\
0 &\lambda
\end{array}
\right]
\left[
\begin{array}{c}
\overline{H}_M \\
\overline{E}_M
\end{array}
\right].
\label{MatrixofHMandEm}
\end{eqnarray}
The eigenvalue and eigenvectors are respectively $\lambda_{1,\, 2}=\pm2\alpha-2(\beta+\eta)$ and $X = \frac{1}{\sqrt{5}}\left[\begin{array}{cc} 2&2\\ 1&-1 \end{array} \right]$. Then,
\begin{eqnarray}
\left[
\begin{array}{cc}
{\overline H}_M(t) \\
{\overline E}_M(t)
\end{array}
\right]=\frac{1}{\sqrt{5}}
\left[
\begin{array}{cc}
2c_1e^{\int^t \lambda_1 d\tau}+2c_2e^{\int^t \lambda_2 d\tau} \\
c_1e^{\int^t \lambda_1 d\tau}-c_2e^{\int^t \lambda_2 d\tau}
\end{array}
\right].
\label{SolutionofHMandEm}
\end{eqnarray}
And, $c_1$ and $c_2$ can be replaced by $H_{M0}$ and $E_{M0}$. We have
\begin{eqnarray}
2\overline{H}_M(t)&=&(2\overline{E}_{M0}+\overline{H}_{M0})e^{2\int^{t}_0(\alpha-\beta-\eta)d\tau}-(2\overline{E}_{M0}-\overline{H}_{M0})e^{2\int^{t}_0(-\alpha-\beta-\eta)d\tau},\label{HmSolutionwithAlphaBeta1}\\
4\overline{E}_{M}(t)&=&(2\overline{E}_{M0}+\overline{H}_{M0})e^{2\int^{t}_0(\alpha-\beta-\eta)d\tau}+(2\overline{E}_{M0}-\overline{H}_{M0})e^{2\int^{t}_0(-\alpha-\beta-\eta)d\tau}.\label{EmSolutionwithAlphaBeta2}
\end{eqnarray}
Realizability condition $2\overline{E}_M>\overline{H}_M$ is certified. Also, for $\alpha<0$, the second terms are dominant leading to the negative $\overline{H}_M$. In contrast, for $\alpha>0$, the first terms are dominant leading to the positive $\overline{H}_M$. Also, in any case, $\overline{E}_M$ is positive. Helicity ratio and sign of polarity in large scale are determined with
\begin{eqnarray}
\overline{f}_{h}=\frac{2\overline{H}_M(t)}{4\overline{E}_{M}(t)}=\frac{(2\overline{E}_{M0}+\overline{H}_{M0})e^{2\int^{t}_0(\alpha-\beta-\eta)d\tau}
-(2\overline{E}_{M0}-\overline{H}_{M0})e^{2\int^{t}_0(-\alpha-\beta-\eta)d\tau}}
{(2\overline{E}_{M0}+\overline{H}_{M0})e^{2\int^{t}_0(\alpha-\beta-\eta)d\tau}
+(2\overline{E}_{M0}-\overline{H}_{M0})e^{2\int^{t}_0(-\alpha-\beta-\eta)d\tau}}\rightarrow \pm1.
\label{fhm_saturated}
\end{eqnarray}
In Fig.~\ref{f3}, when the system is forced with positive kinetic helicity (dash-dotted line, $f_{hk}=1$), the helicity ratio of the large-scale magnetic field $f_{hm}$ (red solid line) decreases and converges to $-1$. Furthermore, when $\overline{E}_M$ begins to rise at $t \sim 150$ (Fig.~\ref{f1}), $f_{hm}$ drops to $-1$ in a staircase-like manner. However, in the case of helical magnetic forcing, the sign of $\alpha$ remains the same as the forcing function leading to $f_{hm}=1$.\\

To find \(\alpha\) and \(\beta\), we can either use direct substitution or apply a simple trick. Here,we introduce the simple method. By multiplying Eq.~(\ref{Em1}) by 2 and subtracting it from Eq.~(\ref{Hm1}), we obtain the equation for \(\overline{H}_M - 2\overline{E}_M\). Conversely, by adding them, we can derive another equation for \(\overline{H}_M + 2\overline{E}_M\). Then, we can proceed to derive the desired expressions:
\begin{eqnarray}
\alpha(t)&=&\frac{1}{4}\frac{d}{dt}log_e \bigg|\frac{ 2\overline{E}_M(t)+\overline{H}_M(t)}{2\overline{E}_M(t)-\overline{H}_M(t)}\bigg|,\label{alphaSolution3}\\
\beta(t)&=&-\frac{1}{4}\frac{d}{dt}log_e\big| \big(2\overline{E}_M(t)-\overline{H}_M(t) \big)\big( 2\overline{E}_M(t)+\overline{H}_M(t)\big)\big|-\eta,\label{betaSolution3}
\label{betaSolution31}
\end{eqnarray}
Compared to conventional approaches, \(\alpha\) and \(\beta\) are functions of only the magnetic helicity and magnetic energy associated with the large-scale magnetic field \(\overline{B}\). Analytically, these representations can be applied to Eqs.~(\ref{Hm1}) and (\ref{Em1}) to yield consistent results. However, we also need to verify them with numerically simulated results before applying them to real data. To obtain profiles, a data set of \(\overline{E}_M(t)\) and \(\overline{H}_M(t)\) from direct numerical simulations is required, with time intervals. We used an approximation such as \(\Delta \overline{E}_M/\Delta t \sim (\overline{E}_M(t_n) - \overline{E}_M(t_{n-1}))/ (t_n - t_{n-1})\). The \(\alpha\) and \(\beta\) profiles shown in Fig.~\ref{f1} and other figures were generated using this method. The IDL script we used is presented in the Appendix.

\subsection{Derivation of $\beta$ with $E_V$ and $H_V$}
Now, we check the possibility of negative $\beta$ using analytic method.
\begin{eqnarray}
&&\langle {\bf u}\times \int^{\tau}(-{\bf u}\cdot \nabla \overline{\bf B})dt \rangle \rightarrow \big\langle -\epsilon_{ijk} u_j (r)u_m(r+l)\tau \frac{\partial \overline{B}_k}{\partial \overline{r}_m}\big\rangle \label{beta_derivation_helical_first}\\
&\sim&-\epsilon_{ijk}\langle u_ju_m\rangle\frac{\partial \overline{B}_k}{\partial \overline{r}_m}-\langle u_j\,l_n\partial_n u_m\rangle\epsilon_{ijk}\frac{\partial \overline{B}_k}{\partial \overline{r}_m}\\
&\sim&\underbrace{-\frac{1}{3}\langle u^2\rangle\epsilon_{ijk}\frac{\partial \overline{B}_k}{\partial \overline{r}_m}\delta_{jm}}_1\,\underbrace{-\big \langle \frac{l}{6}|H_V|\big \rangle\epsilon_{ijk}\frac{\partial
\overline{B}_k}{\partial \overline{r}_m}\delta_{nk}\delta_{mi}}_2,\label{beta_derivation_helical1}
\end{eqnarray}
Here, we set \(\tau \rightarrow 1\) for simplicity, assuming that the two eddies are correlated over one eddy turnover time. We apply a more general identity for the second-order moment as follows \citep{1990cp...book.....M, 2008tufl.book.....L}:
\begin{eqnarray}
U_{jm}\equiv\langle u_j(r)u_m(r+l)\rangle=A(l)\delta_{jm}+B(l)l_jl_m+C(l)\epsilon_{jms}l_s.
\label{general_beta_derivation1}
\end{eqnarray}
With the reference frame of $\vec{l}=(l,\,0,\,0)$ or any other appropriate coordinates, we can easily infer the relation of `$A$', `$B$', and `$C$' as follows: $A+l^2B\equiv F$, $A\equiv G$, $(U_{23}=)lC\equiv H$. Then, Eq.(\ref{general_beta_derivation1}) is represented as
\begin{eqnarray}
U_{jm}=G\,\delta_{jm}+\frac{(F-G)}{l^2}\,l_jl_m+H\epsilon_{jms}\frac{l_s}{l}.
\label{general_beta_derivation2}
\end{eqnarray}
With the incompressibility condition $\nabla \cdot {\bf U}=0$, we get the additional constraint.
\begin{eqnarray}
\frac{\partial U_{jm}}{\partial l_j}=\frac{l_j}{l}G'\delta_{jm}+4l_m\frac{F-G}{l^2}+l_m\frac{(F'-G')l^2-2l(F-G)}{l^3}=0,
\label{constraint_F_G}
\end{eqnarray}
which leads to $G=F+(l/2)\,\partial F/\partial l$. So, the second order moment is
\begin{eqnarray}
U_{jm}=\bigg(F+\frac{l}{2}\frac{\partial F}{\partial l}\bigg)\delta_{jm}-\frac{l}{2l^2}\frac{\partial F}{\partial l}l_jl_m+H\epsilon_{jms}\frac{l_s}{l}.
\label{general_beta_derivation3}
\end{eqnarray}
If $j=m$, $U_{jj}=F=u^2/3 =E_V/6$. And, if $j\neq m$, the relation $\langle \epsilon_{ijk} u_j (r)u_m(r+l)\partial \overline{B}_k/\partial \overline{r}_m\rangle\rightarrow -\langle \epsilon_{ijk}l_jl_m/2l\, \partial F /\partial l \rangle\partial \overline{B}_k/\partial \overline{r}_m$ implies that any `$m$' makes the average negligible. And, for $H$, we introduce Lesieur's approach \citep{2008tufl.book.....L}:
\begin{eqnarray}
H_V&=&\lim_{y\rightarrow x} \bf u(x)\cdot \nabla\times {\bf u(y)}\nonumber\\
&=&\lim_{y\rightarrow x}\epsilon_{ijn} u_i\frac{\partial u_n}{\partial y_j} \nonumber\\
&=&\lim_{l\rightarrow 0}\epsilon_{ijn} \frac{\partial U_{in}(l)}{\partial l_j}\,\,\,(\leftarrow y=x+l)\nonumber\\
&=&\lim_{l\rightarrow 0}\epsilon_{ijn} \epsilon_{ins}\frac{\partial }{\partial l_j}\bigg(H\frac{l_s}{l}\bigg)   \nonumber\\
&=&\lim_{l\rightarrow 0}\epsilon_{ijn} \epsilon_{ins}\bigg(\delta_{js}\frac{H}{l}-\frac{l_jl_s}{l^3}H+\frac{l_jl_s}{l^2}\frac{\partial H}{\partial l}\bigg)\nonumber\\
&=&-\frac{6}{l}H\rightarrow H=-\frac{l}{6}H_V
\label{general_beta_derivation4}
\end{eqnarray}
Then, $U_{jm}$ is
\begin{eqnarray}
U_{jm}=\frac{\langle u^2\rangle}{3}\delta_{jm}-\epsilon_{jms}\frac{l_s}{6}H_V.
\label{general_beta_derivation5}
\end{eqnarray}
EMF by the advection term $-{\bf u}\cdot \nabla \overline{\bf B}$ is
\begin{eqnarray}
\big\langle -\epsilon_{ijk} u_j (r)u_m(r+l)\frac{\partial \overline{B}_k}{\partial \overline{r}_m}\big\rangle&=&
-\frac{1}{3}\langle u^2\rangle \epsilon_{ijk}\frac{\partial \overline{B}_k}{\partial \overline{r}_m}\delta_{jm}+
\epsilon_{ijk}\epsilon_{jms}\frac{l_s}{6}H_V\frac{\partial \overline{B}_k}{\partial \overline{r}_m}\nonumber\\
&\Rightarrow&-\frac{1}{3}\langle u^2\rangle \nabla\times \overline{\bf B}+\frac{l}{6}H_V \big(\nabla \times \overline{\bf B}\big)
\label{general_beta_derivation6}
\end{eqnarray}
For the second term in RHS, we referred to vector identity $\epsilon_{ijk}\epsilon_{jms}=\delta_{km}\delta_{is}-\delta_{ks}\delta_{im}\rightarrow -\delta_{ks}\delta_{im}$ with the consideration of $\nabla\cdot \overline{\bf B}=0$.
\begin{eqnarray}
\big\langle \epsilon_{ijk}\epsilon_{jms}\frac{l_s}{6}H_V\frac{\partial \overline{B}_k}{\partial \overline{r}_m}\big\rangle \rightarrow \langle \epsilon_{jik}\frac{l}{6}H_V\rangle \,\epsilon_{ijk} \frac{\partial \overline{B}_k}{\partial \overline{r}_i}\rightarrow \frac{l}{6}H_V \big(\nabla \times \overline{\bf B}\big)_j.
\label{general_beta_derivation7}
\end{eqnarray}
We used the normal permutation rule and regarded $l_s$ as $l$. Finally,
\begin{eqnarray}
\big( \frac{1}{3}\langle u^2\rangle - \frac{l}{6}H_V \big)\big(-\nabla \times \overline{\bf B}\big)\equiv \beta \big(-\nabla \times \overline{\bf B}\big).
\label{general_beta_derivation8}
\end{eqnarray}
We infer the constraint of `$l$':
\begin{eqnarray}
\langle u(r)u(r+l)\rangle\equiv g(r)\langle u^2(r)\rangle\sim \frac{1}{3}\langle u^2\rangle-\frac{l}{6}\langle {\bf u}\cdot \nabla\times {\bf u}\rangle=\frac{2}{3}E_V-\frac{l}{6}H_V.
\label{new_beta_derivation_helical3}
\end{eqnarray}
For negative $g(r)$, which is a typical property of parallel correlation function \citep{2004Tise.book.....D}, is $l>4E_V/H_V$.\\

\subsection{Revisiting numerical results}
In Fig.~\ref{f5}, we compare $\beta_M$ from $\overline{H}_M$ and $\overline{E}_M$ [Eq.~(\ref{betaSolution3}] with $\beta_V$ from $H_V$ and $E_V$ [Eq.~(\ref{general_beta_derivation8})]. The former uses large-scale magnetic data, while the latter uses small-scale kinetic data. We first tested various small-scale ranges, such as $k=2-4$, $k=2-6$, and $k=2-k_{\text{max}}$. We found that the forcing scale $k=5$ needs to be included. Additionally, the previous condition $l > 4E_V/H_V$  for a negative $\beta$ represents the minimum requirement. With $l \rightarrow \pi$, we obtained a coincident result as the figure demonstrates. The physical meaning of the correlation length $l$ is not yet clear, but this result suggests that it is associated with the wavenumber of the largest eddy scale in the small-scale regime. Additionally, $\beta_V$ performs better than $\beta_M$ in the nonlinear stage when $t > 300$.\\


In Fig.~\ref{f6}, we compared the EMFs using various approaches. The first approach is the EMF from Eq.~(\ref{magnetic_induction_alpha_beta}) (solid line), the second is the EMF derived from $\overline{H}_M$ and $\overline{E}_M$ (semi $\alpha$, semi $\beta$, dotted line), and the third is the EMF calculated using $\alpha$ from $\overline{H}_M$ and $\overline{E}_M$, and $\beta$ from $\overline{H}_V$ and $\overline{E}_V$ (semi $\alpha$, theo $\beta$, dot-dashed line). As briefly mentioned, the data for the large-scale magnetic field $\overline{B}(t)$ can be easily obtained. By considering $\nabla \rightarrow i\,k$ and $k=1$ in Fourier space, $\nabla \times \langle \mathbf{u} \times \mathbf{b} \rangle$ can be indirectly calculated. The other methods are more direct approaches to determining the EMF. The results show that the indirect EMF lies between the two other EMFs. Compared to the real one, semi$-\beta$ is somewhat weak, while theo$-\beta$ is relatively strong. Additionally, the noise-like oscillations for $t > 300$ significantly disrupt the measurement. We believe this is due to the closeness of $\overline{H}_M \sim 2\overline{E}_M$ in Eq.~(\ref{alphaSolution3}) (also refer to Fig.~\ref{f2b}). Although we replaced $\beta$ from the large-scale magnetic data with that from small-scale kinetic data, we still used $\alpha$ derived from $\overline{H}_M$ and $2\overline{E}_M$. We also need to explore an $\alpha$ that uses kinetic data.\\

In Fig.~\ref{f7}, these coefficients were used to reproduce the evolving large-scale magnetic field $\overline{B}(t)$. We utilized the initial seed $\overline{B}(0)$ from the code, along with the arrays for $\alpha$ and $\beta$ (see Appendix).

\section{Summary}
We drove the plasma system with \(\nu = 0.006\) and \(\eta = 0.006\) using helical kinetic energy, collecting data on energy and helicity. With these numerical data, we explored methods to determine \(\alpha\) and \(\beta\) that linearize EMF and the nonlinear dynamo process. Initially, we introduced conventional statistical methods such as MFT, DIA, and EDQNM. We then discussed an alternative approach to finding these coefficients using large-scale magnetic data, specifically \(\overline{H}_M\) and \(\overline{E}_M\) [Eqs.~(\ref{alphaSolution3}) and (\ref{betaSolution3})]. The evolving profiles of \(\alpha\) and \(\beta\) were plotted and compared with the conventional methods. While \(\alpha\) largely agrees with theoretical predictions, \(\beta\), which remains negative, deviates from conventional results. To verify this negative \(\beta\), indicating an inverse cascade of energy via diffusion, and to understand its origin, we derived \(\beta\) using a recursive method and the second moment identity [Eq.~(\ref{general_beta_derivation1})]. This analysis shows that kinetic helicity in the small-scale regime is coupled with large-scale current density, facilitating energy transport toward the large-scale magnetic eddy $-\frac{l}{6}\langle {\bf u}\cdot \nabla\times {\bf u}\rangle \nabla ^2\overline{\mathbf{B}}$.\\

In Figs.~\ref{f4}-\ref{f6}, we compared these coefficients by plotting the EMF. Additionally, in Fig.~\ref{f7}, we reproduced the evolution of the large-scale magnetic field using the calculated \(\alpha\) and \(\beta\) for verification. The reproduced magnetic fields are consistent with the DNS results. Since the two methods for \(\beta\) use different types of data—large-scale magnetic data $\overline{E}_M$ \& $\overline{H}_M$ and small-scale velocity data $E_V$ \& $H_V$, this consistency suggests that the diffusion observed in the plasma has a physical basis in turbulent plasma motions. Typically, diffusion refers to the migration of energy toward smaller scales due to the proportionally decreasing eddy turnover time. However, in a helical field, the toroidal and poloidal components can amplify the magnetic field eddy in a complementary manner, enabling it to propagate toward larger scales, facilitated by the $\alpha$ effect and diffusion.

\begin{figure*}
    {
   \subfigure[$E_{V,\,f}$ \& $E_M$]{
     \includegraphics[width=8.5 cm]{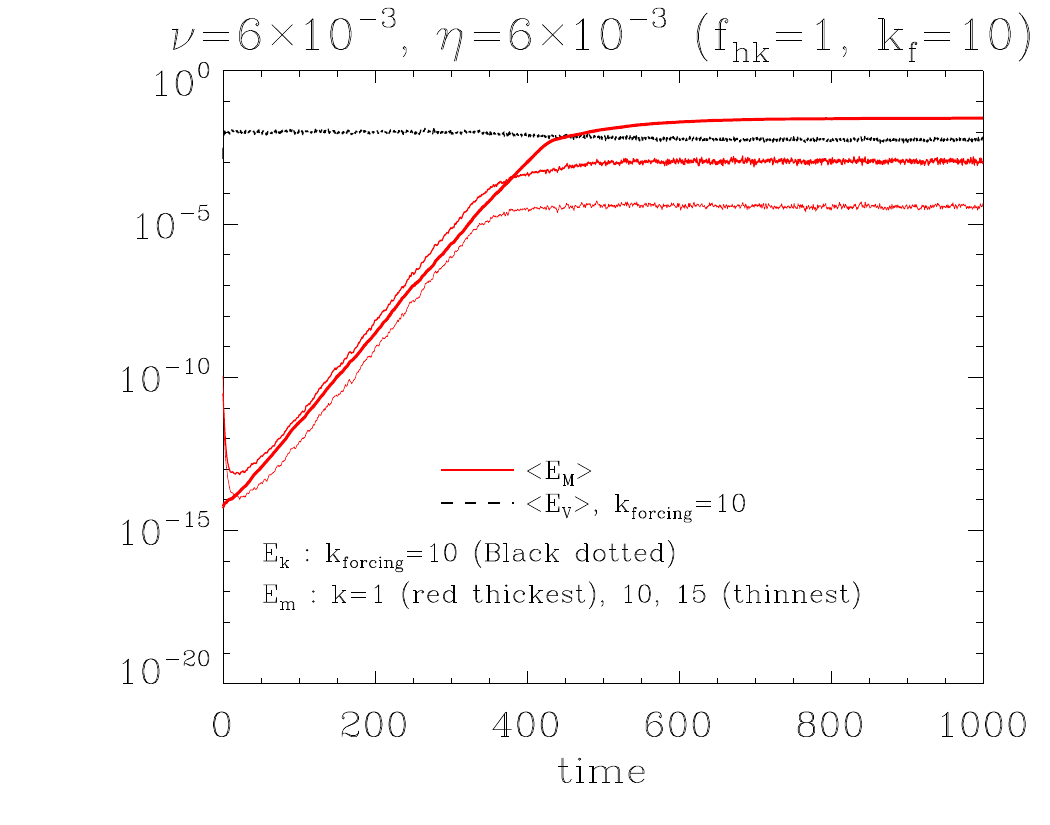}
     \label{f14}   
    }\hspace{-5 mm}
   \subfigure[$E_M$ vs $E_V$]{
   \includegraphics[width=8.5 cm]{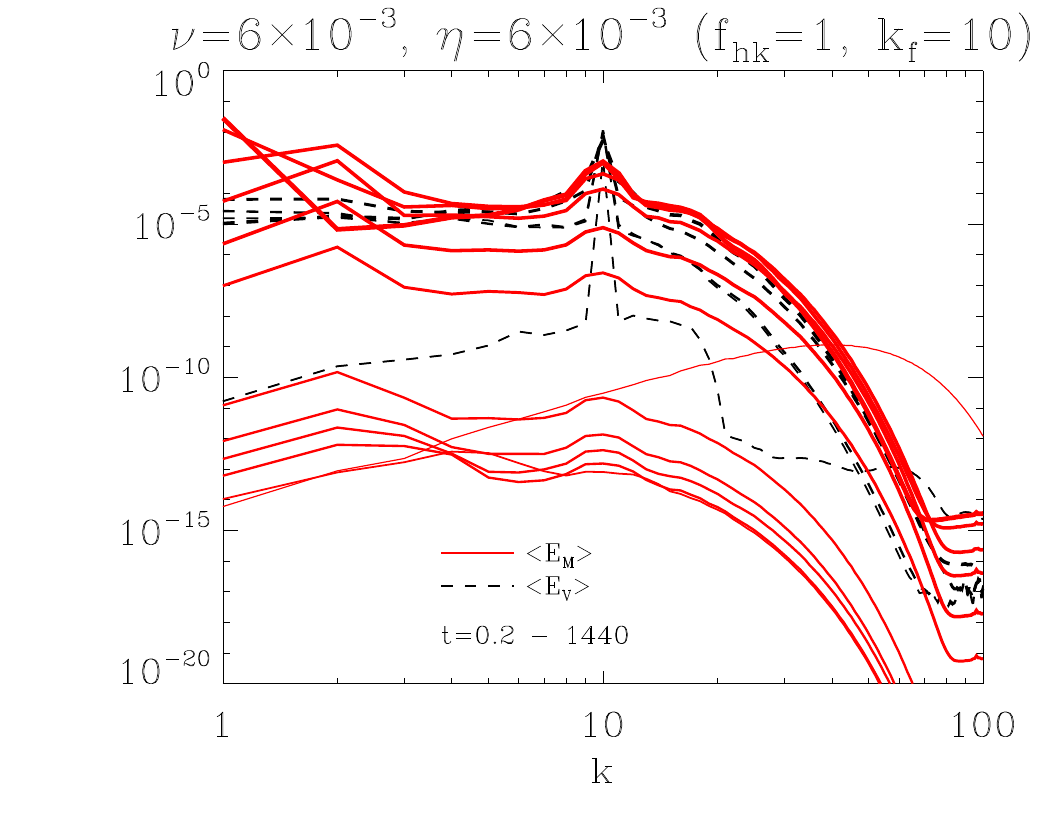}
     \label{f15}  
   }\hspace{-5 mm}
   \subfigure[$\alpha$ vs $\alpha_{MFT}$]{
   \includegraphics[width=8.5 cm]{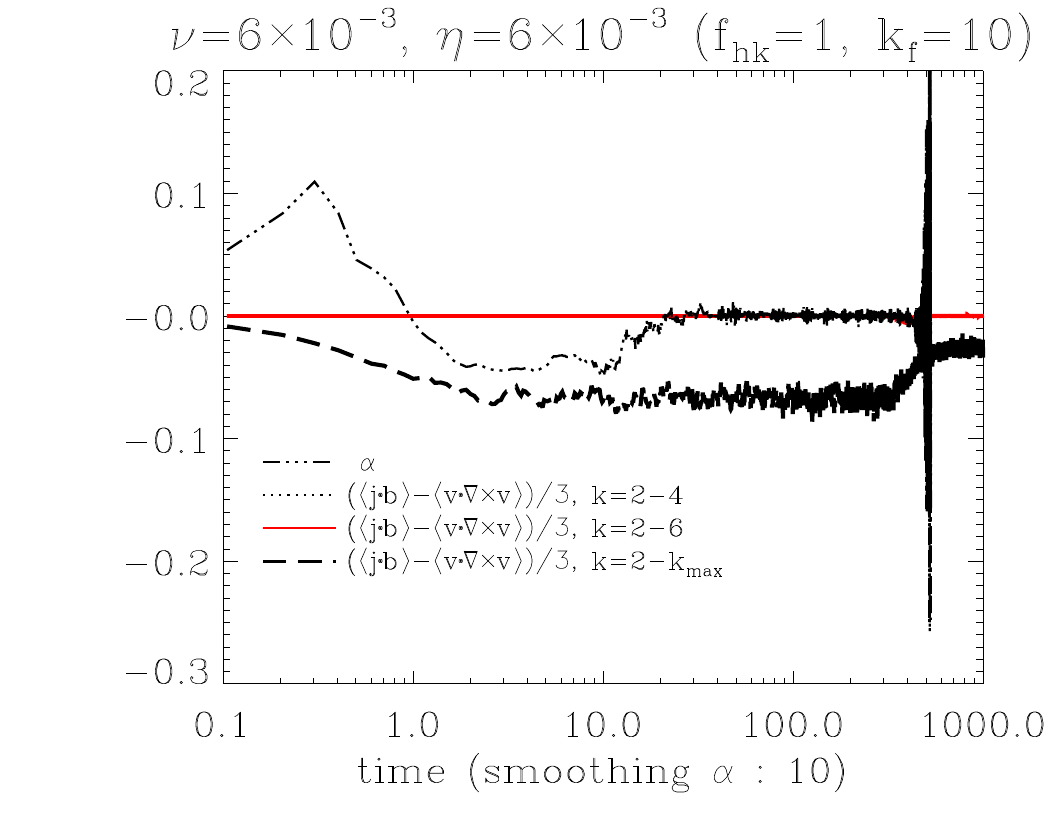}
     \label{f16} 
   }\hspace{-5 mm}
   \subfigure[$\beta$, $\beta_{theo}$, $\beta_{MFT}$]{
     \includegraphics[width=8.5 cm]{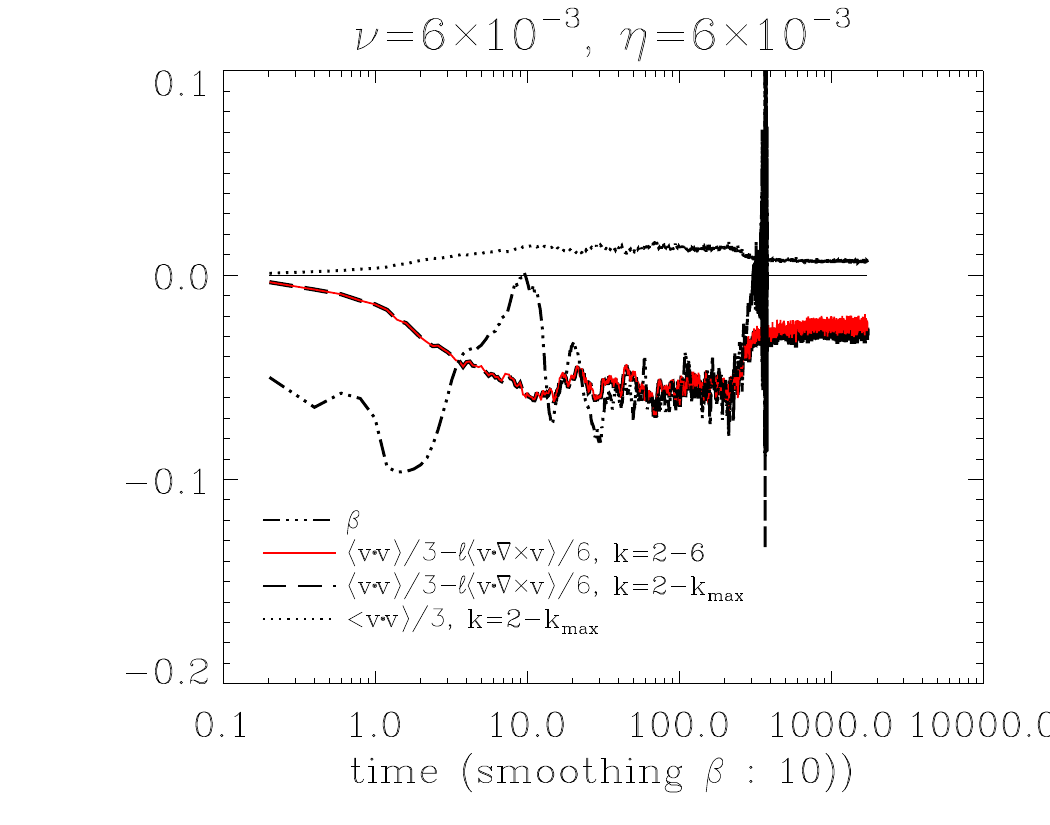} 
     \label{f17}
   }
   \hspace{-5 mm}
   \subfigure[$\nabla \times EMF$]{
     \includegraphics[width=8.5 cm]{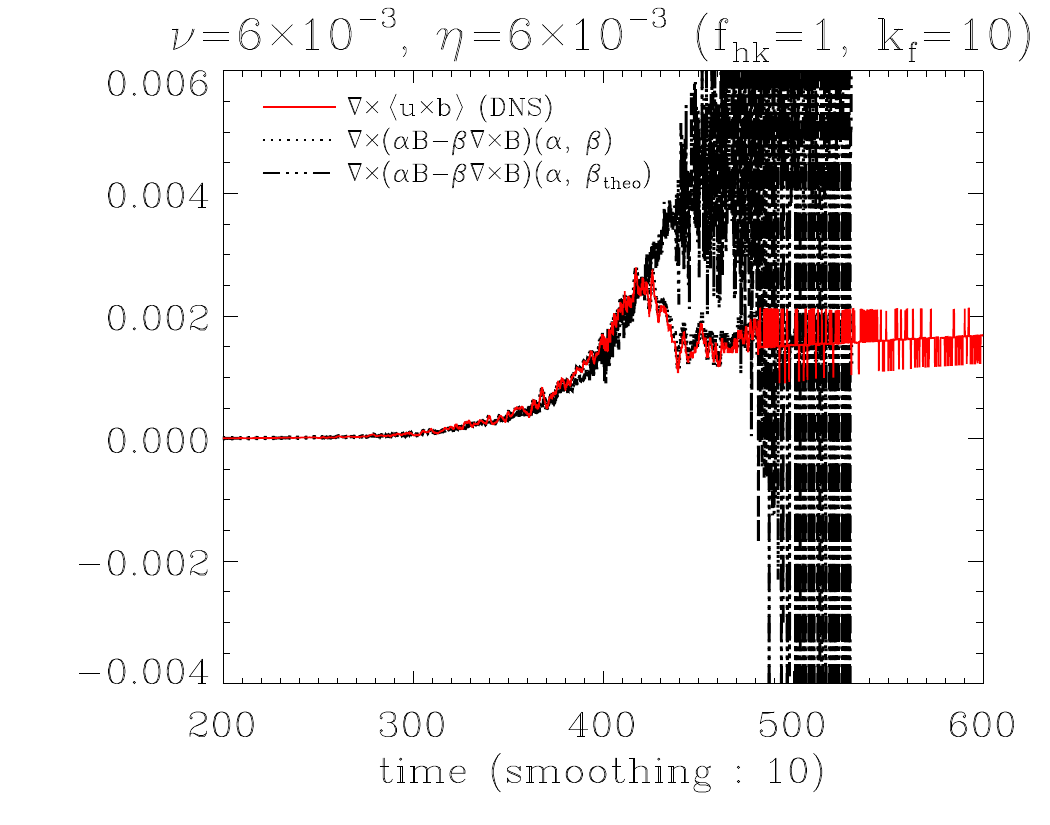} 
     \label{f18}
   }
   \hspace{-5 mm}
   \subfigure[$\langle \overline{B}\rangle$]{
     \includegraphics[width=8.5 cm]{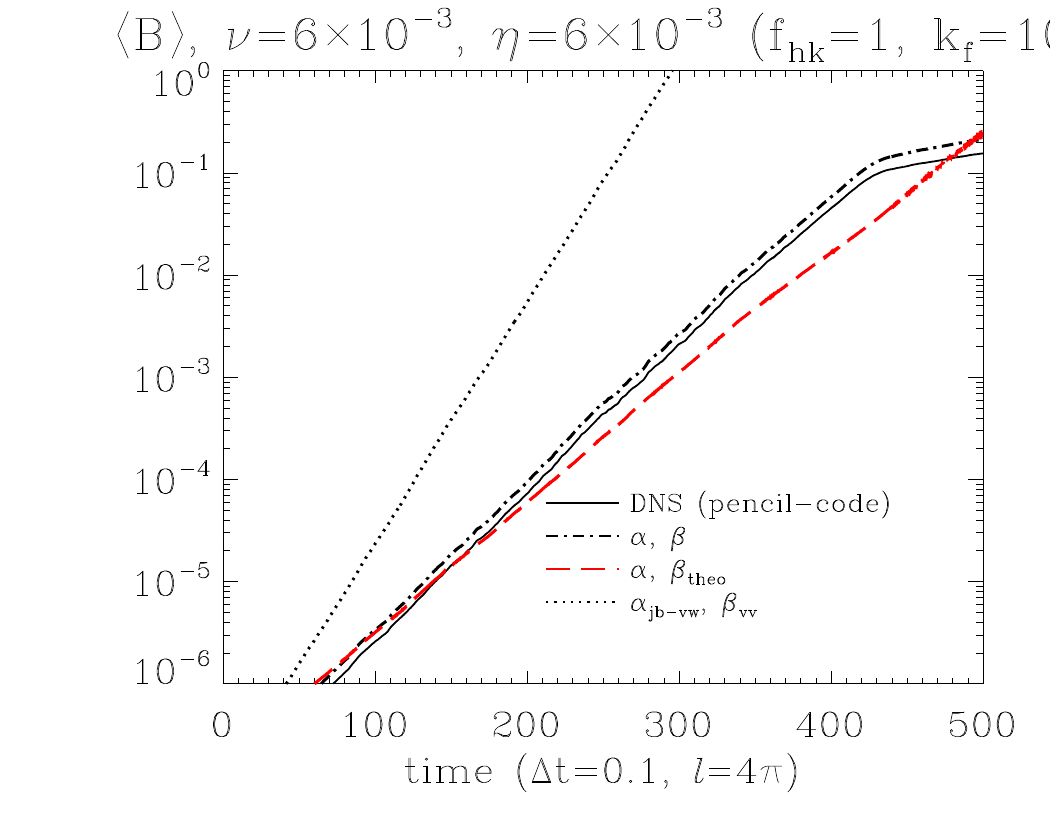} 
     \label{f19}
   }
}
\caption{These figures are of the same type as Figs.~3 and 4, except for the forcing scale at $k=10$.
(a) $\overline{E}_M$ exceeds $E_{V,\,\mathrm{forcing}}$ due to the inverse cascade of converted energy.
(b) Spectra of $E_V$ and $E_M$.
(c) $\alpha$ and $\alpha_{\mathrm{MFT}}$.
(d) $\beta$.
(e) $\nabla \times \mathrm{EMF}$.
(f) $\Delta t = 0.1$, $l = 4\pi$.}
\end{figure*}



\section{Appendix}
Figs.~\ref{f14}-\ref{f19} in the Appendix provide a clearer illustration of the inverse cascade of magnetic energy in the system forced at the smaller scale ($k=10$). These figures are essentially the same type as Figs.~3 and 4. In Fig.~\ref{f14}, the large scale magnetic energy $\langle \overline{E}_M \rangle$ exceeds the forcing energy $\langle E_V \rangle$ at $k=10$, clearly demonstrating the inverse cascade in a helical kinetic forcing dynamo. Fig.~\ref{f14} illustrates the temporal evolution plotted using the spectral data of magnetic energy ($k=1,\,10,\,15$) and kinetic energy ($k=10$). The spectral data of $E_V$ and $E_M$ shown in this figure were utilized in Fig.~\ref{f15}. In Figs.~\ref{f16} and \ref{f17}, $\Delta t = 0.1$ and $l = 4\pi$ were used. The correlation length `$l$` was determined through trial and error. Further detailed research on the correlation length is required. Figs.~\ref{f18} and \ref{f19} verify $\alpha$ and $\beta$. The conventional MFT $\alpha$ and $\beta$ yield a significantly different magnetic field.\\

\noindent The IDL script for Eqs.~(\ref{alphaSolution3}), (\ref{betaSolution3}) is
\begin{verbatim}
for j=0L,  i_last-1 do begin
    c[j]=2.0*spec_mag(1, j) + spechel_mag(1, j)   % k=1 for large scale
    d[j]=2.0*spec_mag(1, j) - spechel_mag(1, j)
endfor

for j=0L,  i_last-1 do begin
  alpha[j]= 0.25*((ALOG(c[j+1])-ALOG(c[j]))-(ALOG(d[j+1])-ALOG(d[j])))/(time[j+1]-time[j])
  beta[j] =-0.25*((ALOG(c[j+1])-ALOG(c[j]))+(ALOG(d[j+1])-ALOG(d[j])))/(time[j+1]-time[j])-eta.
endfor
\end{verbatim}
Here, `spec\_mag', `spechel\_mag' are from pencil\_code power spectrum data for magnetic energy and magnetic helicity in Fourier space.\\

\noindent Also, IDL script for Figs.~\ref{f7}, \ref{f13} is
\begin{verbatim}
B[0] =  sqrt(2.0*spec_mag(1, 0))  % k=1

for j=0L,  t_last do begin
  B[j+1] = B[j] + (-alpha[j]-beta[j]-eta)*B[j]*(time[j+1]-time[j])     % helical magnetic field.
endfor
\end{verbatim}
The negative sign in front of the $\alpha$ coefficient is due to $f_{hm}$ being $-1$ at the large scale $k=1$ (see Figs.~\ref{f4}, \ref{f9}). Except for the initial condition \(\overline{B}(0)\), the evolution of \(\overline{B}\) was self-consistently determined using the separately calculated \(\alpha\) and \(\beta\). The results indicate that these approaches align quite well with each other, and this agreement extends beyond the kinematic regime, which ends much earlier, before they begin to diverge at \(t \sim 300\). Also, $B(t)$ with $\alpha$ \& $\beta$ here and DNS $\overline{B}(t)$ in Figs.~\ref{f6}, \ref{f7} are independently calculated.

\section*{Acknowledgements}
The author acknowledges the support from the physics department at Soongsil University.

\bibliographystyle{apsrev4-2}
\bibliography{bibfile0831} 








\label{lastpage}
\end{document}